\documentclass[10pt]{IEEEtran}
\IEEEoverridecommandlockouts
\usepackage{cite}
\usepackage{amsthm}
\usepackage{amsmath,amssymb,amsfonts}
\usepackage{algorithm}
\usepackage{algorithmicx}
\usepackage{algpseudocode}
\usepackage{amsmath}
\usepackage{mathtools}
\usepackage{graphicx}
\usepackage{textcomp}
\usepackage{xcolor}
\usepackage{bm}
\usepackage{fmtcount}
\usepackage{graphicx,epstopdf,psfrag,url,cite,xcolor,multirow,float}
\usepackage[font=footnotesize]{subfig}

\graphicspath{{figures/}}
\def\BibTeX{{\rm B\kern-.05em{\sc i\kern-.025em b}\kern-.08em
    T\kern-.1667em\lower.7ex\hbox{E}\kern-.125emX}}
\begin{document}
\title{Exploiting NOMA Transmissions in Multi-UAV-assisted Wireless Networks: From Aerial-RIS to Mode-switching UAVs}

\author{Songhan Zhao, Shimin Gong, Bo Gu, Lanhua Li, Bin Lyu, Dinh Thai Hoang, and Changyan Yi \\
\vspace{-1cm}
\thanks{S. Zhao, S. Gong, B. Gu, and L. Li are with the School of Intelligent Systems Engineering, Shenzhen Campus of Sun Yat-sen University, China, and also with the Guangdong Provincial Key Laboratory of Fire Science and Intelligent Emergency Technology, Guangzhou 510006, China (e-mail: zhaosh55@mail2.sysu.edu.cn, \{gongshm5, gubo, lilh65\}@mail.sysu.edu.cn). B. Lyu is with the School of Communications and Information Engineering, Nanjing University of Posts and Telecommunications, China (e-mail: blyu@njupt.edu.cn). D. Hoang is with the School of Electrical and Data Engineering, University of Technology Sydney, Australia (e-mail: hoang.dinh@uts.edu.au). C. Yi is with the College of Computer Science and Technology, Nanjing University of Aeronautics and Astronautics, China (e-mail: changyan.yi@nuaa.edu.cn). }
\thanks{The work of S. Gong was supported in part by National Natural Science Foundation of China under Grant 62372488 and the Shenzhen Fundamental Research Program under Grant JCYJ20220818103201004. The work of L. Li was supported in part by the National Natural Science Foundation of China under Grant 62202506 and the Guangdong University Featured Innovation Program Project (No. 2023KTSCX004).}
}
\maketitle
\thispagestyle{empty}
\begin{abstract}
In this paper, we consider an aerial reconfigurable intelligent surface (ARIS)-assisted wireless network, where multiple unmanned aerial vehicles (UAVs) collect data from ground users (GUs) by using the non-orthogonal multiple access (NOMA) method. The ARIS provides enhanced channel controllability to improve the NOMA transmissions and reduce the co-channel interference among UAVs. We also propose a novel dual-mode switching scheme, where each UAV equipped with both an ARIS and a radio frequency (RF) transceiver can adaptively perform passive reflection or active transmission. We aim to maximize the overall network throughput by jointly optimizing the UAVs' trajectory planning and operating modes, the ARIS's passive beamforming, and the GUs' transmission control strategies. We propose an optimization-driven hierarchical deep reinforcement learning (O-HDRL) method to decompose it into a series of subproblems. Specifically, the multi-agent deep deterministic policy gradient (MADDPG) adjusts the UAVs' trajectory planning and mode switching strategies, while the passive beamforming and transmission control strategies are tackled by the optimization methods. Numerical results reveal that the O-HDRL efficiently improves the learning stability and reward performance compared to the benchmark methods. Meanwhile, the dual-mode switching scheme is verified to achieve a higher throughput performance compared to the fixed ARIS scheme.
\end{abstract}
\vspace{-0.7cm}
\section{Introduction}
The reconfigurable intelligent surface (RIS) has been proposed to improve the transmission performance of future sixth-generation (6G) wireless networks~\cite{Chen-2022}. The RIS is equipped with massive reflecting elements, which can reconfigure the propagation environment by inducing phase shifts on the incident signals. 
The RIS is attracting considerable attention from both academia and industry as a promising solution to improve the network performance in existing wireless networks, such as the RIS-assisted non-orthogonal multiple access (NOMA) networks~\cite{Liu-nomaris2021}. The NOMA technology enables multiple users to share the same spectrum by allocating them into different power domains. Integrating the RIS into the NOMA networks can enhance the channel conditions for more flexible and controllable power adjustments~\cite{Ding-2022pro}. This allows more NOMA users to access and thus achieves a higher transmission performance.

The unmanned aerial vehicle (UAV) as another promising technique is introduced into the RIS-assisted wireless networks to improve the network flexibility and capacity~\cite{Liu-2021}.
The UAV is able to provide the line-of-sight (LoS) connections from air to ground users (GUs), and thus improve the GUs' channel conditions~\cite{Vaezi-2022}. The joint application of the UAV and the RIS has been shown to offer significant advantages to wireless networks~\cite{Liu-2021-sur}. Typically, the RIS can be deployed on the ground to facilitate the air-to-ground channel quality. In particular, deploying the RISs near GUs can help establish virtual LoS links between UAVs and GUs, which allows UAVs to avoid the long-distance flights to bypass obstacles. Besides the terrestrial RIS, the RIS can be also mounted on the UAV becoming the aerial-RIS (ARIS)~\cite{Duo-2023}. 
The ARIS can collaborate with the UAVs to jointly design the trajectories to meet the GUs' dynamic traffic demands and network topology.
As such, the ARIS shows superior adaptability and enhanced reliability in dynamic and complex network environments.
\vspace{-0.3cm}
\subsection{Motivations and Challenges}
In the existing UAV- and ARIS-assisted networks, the UAVs and the ARISs generally keep the fixed roles (e.g., either passive reflection or active radio frequency (RF) communications) during their operations.
However, their advantages will become limited when the wireless channels are severely attenuated by the long transmission distance.
For example, when the ARIS is located far away from the UAVs involving in data transmission, the ARIS needs to approach to the UAVs for channel enhancement via passive reflection. The ARIS's mobility may increase the transmission delay and the UAVs' energy consumption, especially in the large-scale wireless networks. Conversely, when UAVs are in close proximity, they need to tolerate strong interference from each other. Although the NOMA method can be employed to cancel partial GUs' interference by leveraging the channel orthogonality, the performance will be limited when channel differences among GUs become very small. One typical solution to these limitations is to deploy multiple UAVs and ARISs across the wireless network to achieve the widespread coverage. 
However, as the number of UAVs and ARISs increases, it is challenging to coordinate them, especially in a decentralized wireless network.
The fixed-mode UAVs and ARISs may not work efficiently when facing a dynamic and heterogeneous traffic demand over the service area. In particular, when the UAVs' traffic loads become unbalanced, the ARISs have to traverse among different UAVs to achieve the load balance by reshaping the UAVs' channel conditions and capacities. This will result in additional time delay and energy consumption, and thus reduce data transmission efficiency. If each UAV carries the RF transceiver and the passive ARIS simultaneously, it can dynamically adjust the operating mode for either data collection or signal reflection based on the current traffic demands and channel conditions. As such, we expect that the dual-mode UAVs' mode switching scheme can offer a more flexible solution to the challenges mentioned above.

Another challenge lies in the complexity of the optimization problem. 
Generally, the performance maximization of the ARIS-assisted wireless network can be formulated as a joint optimization of the passive beamforming and trajectory planning strategies~\cite{Ge-2023}. These variables are coupled not only with each other but also across different time slots during the entire trajectory. One common approach is to decompose the original problem into multiple subproblems and solve them in an alternative manner~\cite{Zhao-wcnc2024}. The ARIS's passive beamforming and trajectory planning subproblems are often tackled by using the semidefinite relaxation (SDR) and successive convex approximation (SCA) methods, respectively. The approximation and alternating characteristics can obtain a feasible solution but may be far from the optimum of the original optimization problem. 
Besides the optimization methods, the deep reinforcement learning (DRL) provides another feasible approach. The DRL agent continuously adjusts the actions of all UAVs and GUs by interacting with the environment to improve the overall network performance. However, if the agent learns all control variables directly, the high-dimensional action space also leads to an unstable learning performance, making it difficult to learn the meaningful actions. 
\vspace{-0.2cm}
\subsection{Solutions and Contributions}
In this paper, we first explore an ARIS-assisted multi-UAV wireless network, where multiple UAVs serve as aerial access points (APs) to collect the data from the GUs and an ARIS moves around for channel enhancement. To improve the spectral efficiency and support the massive connectivity, we employ NOMA for the GUs' uplink transmissions. Multiple GUs connecting to the same UAV create an NOMA group. The successive interference cancellation (SIC) technique is employed to decode the GUs' signals within each NOMA group. However, the NOMA capacity will become limited if there is limited channel orthogonality in one NOMA group~\cite{Fu-2021TOM}. To tackle this difficulty, the ARIS is used to reshape the channel conditions via passive beamforming and thus induces a desirable channel orthogonality for the GUs' NOMA transmissions. Besides the AIRS's passive beamforming, we also exploit the trajectory planning of both UAVs and ARIS to create the spatial-temporal channel orthogonality for the GUs.

As the ARIS-assisted scheme may become inflexible for a large-scale wireless network, we further develop a novel dual-mode switching scheme aimed at enhancing both channel conditions and data transmission efficiency.
Specifically, the dual-mode switching scheme allows each UAV to carry a passive ARIS and an active antenna system. Moreover, it enables the dual-mode UAVs to switch between the passive and the active modes to perform either the signal reflection or the data transmission on demand. The passive UAVs reshape the air-to-ground channel conditions while the active UAVs serve as aerial APs for data transmissions.
By switching the operating mode on demand, the UAVs can significantly improve the agility to meet the GUs' dynamic traffic demands.
For example, when multiple UAVs are densely deployed, a portion of them can switch to the passive mode to reduce channel competition and signal interference. As the UAVs fly far apart, the interference among them gradually diminishes. This allows the UAVs to switch back to the active mode and thus serve the GUs more efficiently.

We aim to maximize the overall throughput by jointly optimizing the ARIS's passive beamforming, the UAVs' trajectories planning and operating modes, as well as the GUs' transmission control strategies.
We design an optimization-driven hierarchical deep reinforcement learning (O-HDRL) framework, leveraging the benefits of both the optimization and DRL methods. Specifically, the original problem is firstly decomposed into two parts based on the characteristics of the control variables.
The multi-agent deep deterministic policy gradient (MADDPG) algorithm is employed to adjust the trajectories of the UAVs and ARIS. The UAVs' mode switching optimization is a combinatorial problem, which will introduce the significant computational complexity as the number of the UAVs increases. Thus, the UAVs' mode switching strategy is also determined by the MADDPG method. After obtaining the UAVs' trajectories and the mode switching strategies, the optimization methods focus on solving the ARIS's passive beamforming and GUs' transmission strategies. This helps decouple different variables, and thus effectively reducing the complexity. The optimization method as the model-based approach can leverage partial model information from the wireless networks. As such, it can provide supervision to the model-free MADDPG, which is envisioned to further improve the MADDPG's learning performance~\cite{Cui-2023}. The main contributions of this paper are outlined as follows:
\begin{itemize}
\item ARIS-assisted multi-UAV NOMA transmission:
We develop a UAVs' collaborative scheme where multiple UAVs collect data from the GUs using the NOMA method. An ARIS is employed to improve the channel conditions for the GUs' NOMA transmissions by dynamically adjusting its trajectory and passive beamforming strategies. The overall network throughput is maximized by the joint optimization of the ARIS's passive beamforming, the UAVs' trajectories planning, as well as the GUs' transmission strategies.
\item UAVs' mode switching for capacity enhancement:
We develop a dual-mode switching scheme that allows each UAV to dynamically switch between the ARIS mode facilitating other UAVs' transmissions and the aerial AP mode for data transmission from GUs. By providing the UAVs more flexibility, the dual-mode switching scheme can improve the network throughput by $17.51\%$ compared to the fixed-ARIS scheme.
\item Optimization-driven hierarchical DRL algorithm:
We propose a learning algorithm to solve the complex optimization problem efficiently by decoupling the control variables into two parts. The MADDPG method can dynamically adjust the UAVs' trajectory planning and mode switching strategies. The ARIS's passive beamforming and GUs' transmission strategies are optimized by the SDR and SCA methods in each time slot. The proposed algorithm has been shown to reduce computational complexity and improve learning efficiency compared to the conventional learning methods.
\end{itemize}

The rest of this paper is outlined as follows. The related works are discussed in Section~\ref{sec-related}. In Section~\ref{sec-model}, we introduce the fixed-mode ARIS scheme and explain the design ideas behind the O-HDRL framework in Section~\ref{sec-OMADDPG}. To further enhance the system's flexibility and transmission efficiency, we introduce a dual-mode switching scheme in Section~\ref{sec-extension}. Finally, we evaluate the O-HDRL framework and the dual-mode switching scheme in Section~\ref{sec-simulation} and present the conclusions in Section~\ref{sec-conclusion}.
\vspace{-0.3cm}
\section{Related Work}\label{sec-related}
\subsection{RIS-assisted NOMA Transmissions}
Due to the passive characteristics, the RIS is easily employed in the wireless NOMA networks to augment the network capacity.
The authors in~\cite{Zheng-2020} investigated the RIS in both NOMA and orthogonal multiple access (OMA) wireless networks. It is preferable to pair users with asymmetric rates and/or asymmetric deployment to improve the NOMA performance compared to the OMA. The integration of the RIS and NOMA has been shown to reduce energy consumption~\cite{Li-2021con}, improve transmission capacity~\cite{Yang-2021twc}, and enhance flexibility~\cite{Liu-2022wireless}. 
The RIS can also provide an additional channel gain for UAV-assisted wireless networks. The authors in~\cite{Liu-2021} studied the RIS-assisted NOMA transmissions in UAV-assisted wireless networks by jointly optimizing the UAVs' trajectories and the RIS's passive beamforming strategies. The RIS and the UAV can jointly enhance the channel conditions to improve the GUs' NOMA transmission performance. The RIS can also be mounted on the UAVs to provide an extended service coverage for NOMA transmissions. However, deploying an RIS on a UAV can increase its energy consumption, making it challenging to work efficiently in energy-constrained scenarios. To address this, the authors in~\cite{Xiao-2023lett} explored the potential of using the solar energy to extend the ARIS's operational lifespan. The authors in~\cite{Zhang-2023twc} explored an ARIS-assisted wireless NOMA network, where the ARIS can dynamically adjust its trajectory and passive beamforming to meet the GUs' time-varying SIC demands. To further extend the service range, the authors in~\cite{Aung-TVT} investigated a multi-ARIS-assisted wireless network, where multiple ARISs can collaboratively generate passive beamforming to enhance the GUs' channel conditions. The above works demonstrated that the ARIS is capable of offering a more flexible 3D passive beamforming and providing the NOMA GUs with more access opportunities.
However, the fixed-mode ARIS scheme may not perform efficiently in large-scale wireless networks. Since the ARIS's capability of channel enhancement is severely limited by the double fading effect, the ARIS is only effective in a limited coverage range when it is close to either the UAVs or GUs~\cite{Han-2023}.
\vspace{-0.4cm}
\subsection{Multi-UAV Collaborative Data Transmission}
By exploiting UAVs' mobilities, the UAV-assisted wireless networks can effectively overcome many limitations in terrestrial networks. The authors in~\cite{Hua-2020} investigated the UAV-assisted backscatter communication networks, where the UAV operating as the relay assists backscatter users in forwarding signals to the base station. In addition, the UAV can serve as the aerial base station to receive and process computational tasks from the GUs in the mobile edge computing (MEC) system. The authors in~\cite{Hao-2024} employed multiple UAVs in the MEC system, where each UAV either processes the offloaded tasks from the GUs or forwards them to the other UAVs for collaborative computation. The authors in~\cite{Li-2020} optimized the overall throughput in a RIS-assisted UAV network, where the UAV collects data from the GUs with the assistance of the RIS. Besides the data transmission, the authors in~\cite{Wang-2021} investigated the multi-UAV-assisted wireless powered communication (WPC) systems. The multiple UAVs first transfer RF energy to the GUs in the downlink, and then the GUs transmit uplink data to the associated UAVs by using the harvested energy. The multiple UAVs can also play different roles in wireless networks. The authors in~\cite{Xu-2022} studied a dual-UAV cooperative wireless network, where an active UAV equipped with the transceiver transmits control signals to the GUs, while a passive UAV equipped with an ARIS enhances the air-to-ground channel conditions for the active UAV. However, the UAV and the ARIS keep the fixed operating modes during the trajectories, which makes it challenging to tackle the time-varying channel conditions and GUs' traffic demands.
Differently, we focus on the collaborative transmissions and network performance optimization of multiple dual-mode UAVs, which dynamically switch the UAVs' working modes depending on the time-varying network environment.
\vspace{-0.3cm}
\subsection{Performance Maximization in RIS-assisted UAV Networks}
The performance maximization for UAV- and RIS-assisted wireless networks is challenging due to the high dimensionality and deep coupling of different control variables. One typical approach is to simplify the original optimization problem into several subproblems and subsequently solve them individually by the optimization methods. The authors in~\cite{Wu-2020} developed a joint optimization framework for the downlink multi-antenna RIS-assisted system. To minimize the transmit power of the base station, the active and passive beamforming are jointly optimized by using the SDR and alternating optimization (AO). To enhance the interpretability of the solution, the authors in~\cite{Wei-2021} derived a closed-form solution to the sum-rate maximization problem based on the composite channel model in the RIS-assisted UAV systems. In the RIS-assisted WPC systems, the authors in~\cite{Wu-2022} designed different RIS's passive beamforming strategies for the downlink energy transfer and uplink information transmissions, respectively. It was found that setting the same passive beamforming strategy for both energy and information transmission achieves considerable throughput performance while significantly reducing the computational complexity. Compared to the conventional optimization methods, the DRL methods are able to address more complex optimization problems by training the agent via the interaction with the environment~\cite{Feriani-2021}. The authors in~\cite{Wang-2023} designed both deep Q-network (DQN) and deep deterministic policy gradient (DDPG) algorithms for the UAV's trajectory planning. The results show that the DQN requires a shorter training time, while the DDPG has the superior learning performance in terms of reward. Besides the single-agent methods, the multi-agent DRL methods enable the centralized training of multiple agents, which are then distributed to wireless devices for distributed execution.
This allows each device to make independent decisions according to its observation when it is deployed in the large-scale wireless networks. To maximize the overall throughput and energy efficiency, the authors in~\cite{Chen-2023} designed a multi-agent learning approach that combines game theory with proximal policy optimization (PPO). Each agent acts as a game player that independently selects the transmission action based on its local observation. In this paper, we propose an O-HDRL framework combining the advantages of both the optimization and DRL methods. The optimization methods can improve the DRL's learning efficiency and reward performance by offering approximate solutions with partial system information.
\vspace{-0.2cm}
\section{System Model}\label{sec-model}
\begin{figure}[t]
	\centering
	\includegraphics[width = 0.45\textwidth]{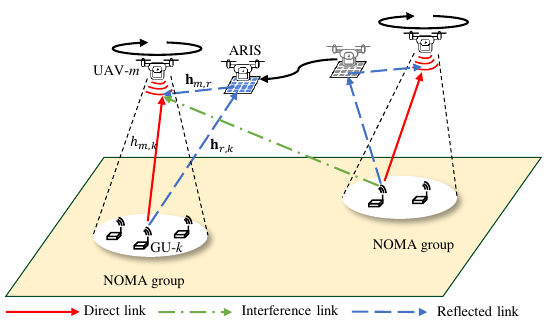}
	\caption{ARIS-assisted multi-UAV wireless networks.}\label{ARIS-struc}
\vspace{-0.6cm}
\end{figure}
As illustrated in Fig.~\ref{ARIS-struc}, we consider an ARIS-assisted multi-UAV wireless network consisting of $M$ UAVs and an ARIS with a set of scattering elements $\mathcal{L}= \{1,\ldots,L\}$. The UAVs, represented by the set $\mathcal{M}=\{1,\ldots,M\}$, operate as mobile APs to collect and process data from the GUs, while the ARIS moves around to enhance channel conditions for the UAVs' data transmissions. Let $\mathcal{K}=\{1,\ldots,K\}$ represent the set of the GUs. We denote the $m$-th UAV and $k$-th GU as the UAV-$m$ and GU-$k$, respectively. Each GU gathers sensing data from its surroundings and then transmits it to the UAVs.
We consider that all UAVs and GUs are equipped with a single antenna and share the same frequency band. To improve the spectral efficiency, the GUs employ NOMA to transmit data to UAVs.
The GUs connected to the same UAV form an NOMA group. By employing the NOMA, all UAVs can efficiently multiplex the wireless resources while tolerating a certain level of interference. However,  simply increasing the number of UAVs may not improve the data collection performance, as it can increase the co-channel interference among the UAVs and introduce excessive deployment cost. Therefore, to improve the system's transmission efficiency, we employ a low-cost ARIS to offer more flexible channel enhancements that adapts to the dynamic network conditions.
\vspace{-0.4cm}
\subsection{ARIS-enhanced Channel Model}
The location of the GU-$k$ is represented by the two-dimensional coordinate ${\bf z}_k$. All the UAVs and the ARIS work at the fixed altitude $H$. We consider a time-slotted frame structure for the UAVs' operations and let $\mathcal{N}=\{1,\ldots,N\}$ denote the set of all time slots with a unit duration $\tau$.
In the $n$-th time slot, the location of the UAV-$m$ and the ARIS are denoted by ${\bf q}_m[n]$ and ${\bf q}_r[n]$, respectively. During the trajectory, all the UAVs and the ARIS need to ensure the safety distance between each other and satisfy the speed limit, which leads to the following mobility constraints:
\begin{subequations}\label{mobility_safety}
\begin{align}
&d_{i}[n] \triangleq \|{\bf q}_i[n]-\mathbf{q}_i[n-1]\|\leq \tau V_{\mathrm{max}},  \label{trajectory_1}\\
&d_{i,i'}[n] \triangleq \|{\mathbf q}_i[n]-{\mathbf q}_{i'}[n]\|\geq D_{\mathrm{min}},\label{trajectory_2}
\end{align}
\end{subequations}
where $i\neq i'$ and $i,i'\in \mathcal{I} = \{1,\ldots,M+1\}$ represent the indexes of the UAVs and the ARIS. The distance of each UAV (or ARIS) traveled in adjacent time slots is represented by $d_{i}[n]$ and the distances among all the UAVs and the ARIS are represented by $d_{i,i'}[n]$. The constants $ V_{\mathrm{max}}$ and $D_{\mathrm{min}}$ denote the maximum speed and safety distance, respectively.

The channels from the UAV-$m$ to the ARIS, the ARIS to the GU-$k$, and the UAV-$m$ to the GU-$k$ are denoted by ${\bf h}_{m,r}[n]=\beta^{1/2} d^{-\alpha_0/2}_{m,r}[n]{\bf a}_{m,r}$, ${\bf h}_{r,k}[n]=\beta^{1/2} d^{-\alpha_0/2}_{r,k}[n]{\bf a}_{r,k}$ and ${h}_{m,k}[n]=\beta^{1/2} d^{-\alpha_0/2}_{m,k}[n]{a}_{m,k}$, respectively, where $d_{i,k}$ denotes the distance between the UAVs (or ARIS) and the GUs. The constant $\alpha_0$ denotes the path loss exponent and $\beta$ is the channel power gain at a reference distance of one meter. The antenna array response coefficients are denoted as ${\bf a}_{m,r}$, ${\bf a}_{r,k}$, and $a_{m,k}$, respectively~\cite{Khalili-2022}. The ARIS is capable of adjusting the phase-shift of air-to-ground links, enabling channel enhancement and interference mitigation. The phase-shift vector of the ARIS is denoted as $\bm{\theta}[n]=[e^{j\phi_{l}}[n],\ldots,e^{j\phi_{L}}[n]]^T$, where $e^{j\phi_{l}}[n]$ is the phase-shift induced by the ARIS's $l$-th scattering element. In the following, we omit the time index of each variable for brevity. The equivalent channel from the GU-$k$ to the UAV-$m$ is represented as:
\begin{equation}\label{equ_channel}
\widehat{{\bf h}}_{m,k}={\bf h}_{m,r}\text{diag}({\bf h}_{r,k})\bm{\theta}+h_{m,k}\triangleq{\bf H}_{m,k}\bm{\theta}+h_{m,k},
\end{equation}
where ${\bf H}_{m,k}={\bf h}_{m,r}\text{diag}({\bf h}_{m,k})$ and $\text{diag}({\bf h}_{m,k})$ denotes the diagonal matrix with the diagonal  vector ${\bf h}_{m,k}$. The equivalent channel in \eqref{equ_channel} can be dynamically controlled by the ARIS's passive beamforming and trajectory planning strategies.
\vspace{-0.3cm}
\subsection{UAV-assisted NOMA Transmissions}
We consider that all GUs employ NOMA technique to transmit data to the UAVs, which is expected to achieve a higher transmission capacity than that of the conventional OMA schemes~\cite{Ding-2022twc}.
Note that the UAVs are not required to simultaneously connect all GUs, in each time slot, each UAV adaptively selects a subset of GUs for the NOMA transmissions depending on the GUs' channel conditions and  traffic demands.
Let $\rho_{m,k}$ denote the binary variable and $\rho_{m,k}=1$ represents the association between the GU-$k$ and UAV-$m$.
As each GU can transmit data to at most one UAV in each time slot, we have the following association constraint:
\begin{equation}\label{association}
\sum_{m\in\mathcal{M}}\rho_{m,k}\leq1 \text{ and } \rho_{m,k}\in\{0,1\}, \forall m\in\mathcal{M},\forall k\in\mathcal{K}.
\end{equation}

The interference to the GU-$k$ contains the intra- and inter-UAV interference, respectively. The intra-UAV interference $I_{m,k}$ denotes the interference to GU-$k$ caused by the other GUs associated with the same UAV-$m$. The inter-UAV interference $I'_m$ represents the interference introduced by the GUs associated with the other UAVs.
The SIC technique allows the UAVs to mitigate the intra-UAV interference by decoding and canceling the signals one by one. 

The decoding order of the GUs can be determined by the GUs' channel conditions to the UAVs. However, the channel conditions are not only related to the distance between the UAVs and the GUs but also correlated with the ARIS's passive beamforming strategy $\bm{\theta}$. Intuitively, we can formulate a joint optimization problem for the ARIS's passive beamforming and the NOMA decoding order. However, this can introduce additional computational overhead, which potentially hinders the real-time control of the UAVs and the ARIS. Instead, we propose a heuristic method to order the GUs' information decoding. We first consider an optimistic case in which the ARIS is capable of aligning all channels between GUs and the associated UAVs. As such,  the channels between the GUs and the UAVs (denoted by $\widehat{{\bf h}}'_{m,k}$) are calculated as follows:
\begin{equation}\label{optimistic_channel}
\Vert\widehat{{\bf h}}'_{m,k}\Vert =\sum_{l\in\mathcal{L}}\vert { h}_{m,r,l} { h}_{r,k,l}\vert+ \vert h_{m,k} \vert,
\end{equation}
where ${ h}_{m,r,l}$ and $ { h}_{r,k,l}$ are the $l$-th elements of ${\bf h}_{m,r}$ and $ {\bf h}_{r,k}$, respectively.
Let $\mathcal{K}^m = \{1,\ldots,D_m\}$ denote the set of the GUs associated with the UAV-$m$, where $D_m$ is the number of associated GUs. Note that this method effectively mitigates the impact of the ARIS's passive beamforming on channel gains by assuming the upper bound of ARIS's reflection capability.

Then, the GUs' decoding order is determined by the sorting of their channel strengths as follows:
\begin{equation}\label{channle_order}
\Vert\widehat{{\bf h}}'_{m,1}\Vert^2\!\geq\!\ldots\!\geq\Vert\widehat{{\bf h}}'_{m,d}\Vert^2\geq\!\ldots\!\geq\Vert\widehat{{\bf h}}'_{m,D_m}\Vert^2,\! \forall d\in\mathcal{K}^m.
\end{equation}
Following the ordering rule in \eqref{channle_order}, we denote $s_m(k)\in\mathcal{K}^m$ as the decoding order of the GU-$k$ when it is associated with the UAV-$m$. According to the SIC principle, the intra-UAV interference to the GU-$k$ is represented as follows:
\begin{equation}\label{intra_interfernece}
I_{m,k}= \sum\limits_{d > s_m(k),d \in \mathcal{K}^m }\rho_{m,d}p_G\Vert \widehat{{\bf h}}_{m,d}\Vert^2,
\end{equation}
where $p_G$ is the constant GUs' transmit power. The inter-UAV interference to the  UAV-$m$ is calculated as:
\begin{equation}\label{inter_interfernece}
I'_{m}=\sum\limits_{m'\neq m,m'\in\mathcal{M}}\sum\limits_{k\in\mathcal{K}}\rho_{m',k}p_G\Vert \widehat{{\bf h}}_{m,k}\Vert^2.
\end{equation}

Building on \eqref{intra_interfernece} and \eqref{inter_interfernece}, the signal-to-interference-plus-noise ratio (SINR) for the GU-$k$ is expressed as:
\begin{align}\label{SINR-N}
\text{SINR}_{m,k}=\frac{\rho_{m,k}p_G\Vert \widehat{{\bf h}}_{m,k}\Vert^2}{ I_{m,k}+I'_{m}+\sigma^2},
\end{align}
where $\sigma^2$ denotes the background noise power. For the last decoded GU, its SINR calculation removes all intra-UAV interference and only suffers from the inter-UAV interference.
To ensure successful decoding, we have the following minimum SINR requirement constraint in each time slot:
\begin{equation}\label{threshold}
\text{SINR}_{m,k} \ge \gamma, \forall m \in \mathcal{M}, \forall k \in \mathcal{K},
\end{equation}
where $\gamma$ denotes the decoding threshold of the SIC receivers. To this end, the data rate from the GU-$k$ to the UAV-$m$ is represented by $r_{m,k} = \log(1+\text{SINR}_{m,k})$.

In each time slot, all GUs consistently perform sensing the surrounding information. Each GU will offload the sensing data to one UAV when they establish an association. Let $D_k[n]$ denote the GU-$k$'s data buffer size in the $n$-th time slot and $D_{\max}$ represent the maximum capacity of GUs' data queue. Thus, we can represent the GU-$k$'s buffer dynamics as follows:
\begin{equation}\label{data_queue}
D_k[n]\!=\!\left[\bigg[D_{k}[n-1]+D_{s,k}[n]\bigg]^{\!D_{\max}}\!\!\!\!\!\!\!\!-\sum_{m\in\mathcal{M}}\!\tau r_{m,k}[n]\right]_0\!\!,
\end{equation}
where $[\cdot]^{D_{\max}}\triangleq \min\{\cdot,D_{\max}\}$ and $[\cdot]_0\triangleq\max\{\cdot,0\}$. The GUs' sensing data at the beginning of each time slot has a random size, which is denoted as $D_{s,k}[n]$. The collected data by the UAV-$m$ in each time slot is represented as follows:
\begin{equation}\label{collected_data}
R_m=\sum_{k\in\mathcal{K}}\min\{\tau r_{m,k},D_k\}.
\end{equation}
Note that the UAVs' trajectories and the ARIS's passive beamforming dynamically control the channel conditions for data transmission. Thus, we can jointly optimize them to improve the GUs' transmission performance.
\section{Optimization-driven Learning Solution to Throughput Maximization Problem}\label{sec-OMADDPG}
We aim to maximize the overall network throughput by jointly optimizing all UAVs' trajectories  $\{{\bf q}_i\}_{i\in\mathcal{I}}$, GUs' association $\bm{\rho}=\{\rho_{m,k}\}_{m\in\mathcal{M}, k\in\mathcal{K}}$, and the ARIS's passive beamforming $\bm{\theta}$ strategies. The throughput maximization problem is formulated as follows:
\begin{equation}\label{ori_question}
\max_{{\bf q}_i,\bm{\rho},\bm{\theta}}\sum_{n\in\mathcal{N}}\sum_{m\in\mathcal{M}}R_m~~
\mathrm {s.t.}~~\eqref{mobility_safety}-\eqref{collected_data}.
\end{equation}
The ARIS's passive beamforming dynamically reconfigures the channel conditions and thus influences the UAVs' decisions on their next positions. Meanwhile, the UAVs' trajectories determine the GUs' data transmissions. 
Typically, it is possible to use the classic AO method to tackle problem~\eqref{ori_question}. However, it generally needs a central server to collect the global information to perform the optimizations, and then distribute the solutions to each wireless device, which will cause the huge signaling and computational overhead. Moreover, executing the optimization ahead of the UAVs' flight process relies on the future channel state information (CSI), which is difficult to obtain in dynamic network environments. 
Another feasible approach is to use the DRL method that is more flexible to handle the high-dimensional and complex problems. A straightforward application of the DRL method to solve problem~\eqref{ori_question} is to establish multiple agents to jointly adapt the UAVs' trajectories, the GUs' association, and the ARIS's passive beamforming strategies. However, the high-dimensional action space requires a large number of training samples to attain sufficient information for mature actions.

\begin{figure}[t]
	\centering
	\includegraphics[width = 0.45\textwidth]{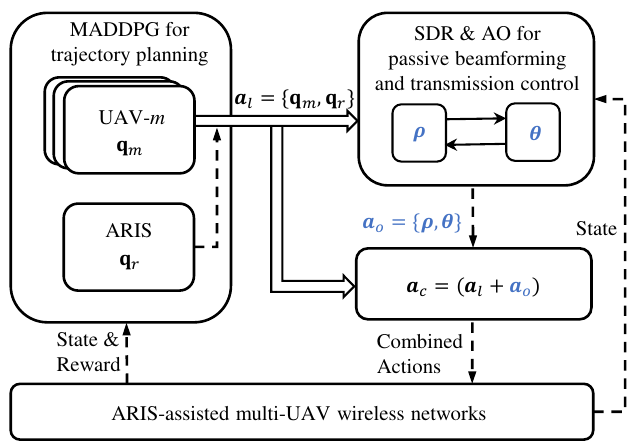}
	\caption{The proposed O-HDRL framework.}\label{diagram}
\vspace{-0.5cm}
\end{figure}

The intuition of the proposed O-HDRL framework is to decompose problem~\eqref{ori_question} into two parts, as shown in Fig.~\ref{diagram}. The MADDPG agents first adjust all UAVs' next positions (denoted as $\bf{q}_m$ and $\bf{q}_r$ for the UAV-$m$ and the ARIS, respectively) by fixing the ARIS's passive beamforming $\bm{\theta}$ and GUs' association $\bm{\rho}$. After obtaining the UAVs' trajectories, the ARIS's passive beamforming and GUs' association strategies are adapted by the optimization methods. The multiple agents execute the combined action from both DRL and optimization in the wireless environment. The state is updated and the feedback reward guides the agents' next round of learning. The optimization part can be also regarded as supervision to drive the DRL's learning process~\cite{Zhao-2023}. The above concept is expected to efficiently improve the DRL training rate and learning performance. Additionally, the UAVs can make online decisions based on the real-time CSI in each time slot without the need for the future CSI, which makes it more suitable for practical implementation.
The MADDPG method can be replaced with other MADRL algorithms as needed. We focus on integrating the model-based optimization module into the model-free MADRL method to improve their learning efficiency and performance.
In the sequel, we detail the O-HDRL framework for optimization and learning parts, respectively.
\vspace{-0.3cm}
\subsection{MADDPG for Trajectory Planning}
We first reformulate the UAVs' trajectory planning optimization into a Markov decision process (MDP), which is characterized as a tuple ($\mathcal{S}, (\mathcal{A}_{l},\mathcal{A}_{o}),\mathcal{R}$). The state space is denoted as $\mathcal{S}$, where $\mathbf{s} = \{\mathbf{s}_i\}_{i\in \mathcal{I}}\in\mathcal{S}$ denote the UAVs' states including the channel information and the UAVs' locations. The action space is divided into two subsets, i.e., the learning action space $\mathcal{A}_{l}$ and the optimization action space $\mathcal{A}_{o}$. The learning action $\mathbf{a}_l=\{\mathbf{a}_{l,i}\}_{i\in\mathcal{I}}\in\mathcal{A}_{l}$ is adapted by the MADDPG, which includes the UAVs' flying direction and speed in each time slot. The optimization action $\mathbf{a}_{o}\in\mathcal{A}_{o}$ consists of the ARIS's passive beamforming and GUs' association strategies obtained by the optimization method. We define the UAVs' combined actions from the MADDPG and the optimization as $\mathbf{a}_c=\{\mathbf{a}_{c,i}\}_{i\in\mathcal{I}}$ with $\mathbf{a}_{c,i} = \{\mathbf{a}_{l,i},\mathbf{a}_{o}\},i\in\mathcal{I}$. The reward $R(\mathbf{s}_i, \mathbf{a}_{c,i})\in\mathcal{R}$ evaluates the goodness of the state-action pair $(\mathbf{s}_i, \mathbf{a}_{c,i})$, and then drives each agent to make better decisions.

We design distinct reward functions for the agents of the UAVs and the ARIS, as they have the different objectives during their operations. In particular, the UAVs focus on maximizing data transmission from the GUs, while the ARIS aims to improve the channel conditions for the UAVs. Thus, we define the reward functions $R_{m,u}$ and $R_{r}$ for the UAV-$m$ and the ARIS, respectively, as follows:
\begin{subequations}\label{reward_FIXARIS}
\begin{align}
&R_{m,u}=\ell_u\!\sum_{k\in\mathcal{K}^m}\!\min\{\tau\log(1+\text{SINR}_{m,k}),D_k\}-\jmath_u,  \label{rewards_1}\\
&R_{r}=\ell_r\sum_{m \in \mathcal{M}}\sum_{k\in\mathcal{K}}\rho_{m,k}{\bf h}_{m,r}\text{diag}({\bf h}_{r,k})\bm{\theta}-\jmath_r,\label{rewards_2}
\end{align}
\end{subequations}
where $\ell_u$ and $\ell_r$ are the weighting coefficients of the UAV-$m$ and the ARIS, respectively. The first part of $R_{m,u}$ denotes the UAV-$m$'s collected data from the GUs. The first part of $R_{r}$ represents the ARIS's channel improvement to all UAVs. The factors $\jmath_u$ and $\jmath_r$ are the penalty terms assigned to the UAVs and the ARIS, respectively, if the constraints in problem~\eqref{ori_question} do not hold. The MADDPG learning process continues until all agents's policies yield the stable rewards.

During the learning, each agent possesses a pair of actor and critic networks. These networks serve to approximate the policy and value function enabling the agents to adapt and evaluate actions within each training step~\cite{Luong-2019SUR}. The weighting parameters of the $i$-th agent's actor and critic networks are denoted as $\bm{\phi}_i$ and $\bm{\omega}_i$, respectively. Given the current state $\mathbf{s}_i$, the actor network $\bm{\pi}(\mathbf{s}_i|\bm{\phi}_i)$ generates a deterministic policy $\mathbf{a}_{l,i} = \bm{\pi}(\mathbf{s}_i|\bm{\phi}_i)$ to maximize the value function $\mathcal{J}_i({\bm\phi}_i)$, which is evaluated by the $i$-th critic networks defined as follows:
\begin{equation}\label{value-function}
\mathcal{J}_i({\bm\phi}_i) \approx \mathbb{E}_{ \mathcal{D}}\left[Q_i(\mathbf{s},\mathbf{a}_{c,i}|{\bm\omega}_i)\right].
\end{equation}
The Q-value $Q_i(\mathbf{s},\mathbf{a}_{c,i}|{\bm\omega}_i)$ with weighting parameter ${\bm\omega}_i$ is generated by the $i$-th critic network to evaluate the performance of state-action pair $(\mathbf{s},\mathbf{a}_{c,i})$. Given the samples from the experience replay buffer $\mathcal{D}$, the expectation of the Q-value approximates $\mathcal{J}_i({\bm\phi}_i)$, which depends on both $\bm{\phi}_i$ and ${\bm\omega}_i$ from the $i$-th actor and critic networks, respectively.
We update the weighting parameter ${\bm\phi}_i$ of each actor network to enable the agents to choose more preferable actions.
The actor network aims to maximize the value function $\mathcal{J}_i({\bm\phi}_i)$ evaluated from the critic network. Hence, we update the actor network's parameter $\bm{\phi}_i$ by using the deterministic policy gradient method~\cite{Arulkumaran2017}, and the derivative of $\mathcal{J}_i({\bm\phi}_i)$ is represented as follows:
\begin{equation}
\nabla\mathcal{J}_i({\bm\phi}_i) = \mathbb{E}_{ \mathcal{D}}\left[\nabla_{\mathbf{a}_{l,i}}[Q_i(\mathbf{s},\mathbf{a}_{c,i}|{\bm\omega}_i)\nabla_{{\bm\phi}_{i}}\bm{\pi}(\mathbf{s}_i)\right].
\end{equation}
The $i$-th critic network is updated by the temporal difference (TD) error. The TD error is computed as the difference between the real-time Q-value $Q_i(\mathbf{s},\mathbf{a}_{c,i}|{\bm\omega}_i)$ and its target value $y_t = R(\mathbf{s}_i, \mathbf{a}_{c,i}) +\gamma_d Q'_i$, where $ Q'_i$ denotes the Q-value estimated from the target critic network. The discount factor $\gamma_d$ influences the agents' preference between the short-term and long-term rewards. To minimize the TD error, the loss function $L({\bm\omega}_i)$ of the $i$-th critic network is defined as follows:
\begin{equation}
L({\bm\omega}_i) = \mathbb{E}_{\mathcal{D}}[(y_t-Q_i(\mathbf{s},\mathbf{a}_{c,i}|{\bm\omega}_i))^2].
\end{equation}
By adjusting the weighting parameter $\bm{\omega}_i$ for each critic network, it provides an increasingly accurate estimation of the Q-value to evaluate the agents' decision making.
The iterative training of the actor and critic networks drives agents to take action with better reward in dynamic wireless environment.
\vspace{-0.3cm}
\subsection{Passive Beamforming and Transmission Control}\label{subproblem-opti}
Given the UAVs' trajectories, we focus on the ARIS's passive beamforming and GUs' association strategies, which are optimized by solving the following subproblem:
 \begin{equation}\label{sub_question}
\max_{\bm{\rho},\bm{\theta}}\sum_{n\in\mathcal{N}}\sum_{m\in\mathcal{M}}R_m~~
\mathrm {s.t.}~~\eqref{equ_channel}-\eqref{collected_data}.
\end{equation}
Problem~\eqref{sub_question} is challenging to solve due to the coupling feature between $\bm{\rho}$ and $\bm{\theta}$. Hence, we propose an AO method to decompose problem~\eqref{sub_question} into two subproblems (i.e., ARIS's passive beamforming and GUs' transmission control subproblems) and solve them alternately.

We first solve the ARIS's passive beamforming subproblem and reformulate the UAV-$m$'s rate $R_m^l$ as follows:
\begin{align}
R_m^l
&\!\!=\log\Big(\sum_{d\in\mathcal{K}^m} \rho_{m,d}p_G\text{Tr}({\bf Q}_{m,d} {\bf\Phi}) \notag\\
&\!\!+ \sum_{m'\neq m,m'\in\mathcal{M}}\sum_{d\in\mathcal{K}^{m'}}\rho_{m',d}p_G\text{Tr}({\bf Q}_{m,d} {\bf\Phi})+\sigma^2\Big)\notag\\
&\!\!-\log\Big(\!\!\!\sum_{m'\neq m,m'\in\mathcal{M}}\!\sum_{d\in\mathcal{K}^{m'}}\!\!\rho_{m',d}p_G\text{Tr}({\bf Q}_{m,d} {\bf\Phi})\!+\!\sigma^2\Big)\label{second_term}.
\end{align}
The auxiliary variables $\{{\bf Q}_{m,d}\}_{m\in\mathcal{M},d\in \mathcal{K}}$ and ${\bf\Phi}$ introduced into \eqref{second_term} are represented as follows:
\begin{subequations}\label{intro_var}
\begin{align}
&{\bf Q}_{m,d}=\begin{bmatrix} ~{\bf H}^{H}_{m,d}{\bf H}_{m,d} &{\bf H}^{H}_{m,d} h_{m,d} \\ ~h^{H}_{m,d}{\bf H}_{m,d} &h^{H}_{m,d}h_{m,d} \\ \end{bmatrix},\label{q_matrix}\\
&{\bf\Phi}=[\bm{\theta};1][\bm{\theta};1]^H \succ 0 \text{ and } {\Phi}_{l,l} = 1,~l=1,\!\ldots\!,L+1,\label{beam_matrix}
\end{align}
\end{subequations}
where ${\Phi}_{l,l}$ denotes the diagonal element of ${\bf\Phi}$. The second term in~\eqref{second_term} can be linearly approximated by introducing auxiliary variables $\{\ell_{m}\}_{m\in\mathcal{M}}$ given point $\ell_{m,0}$, which is reformulated as follows:
\begin{subequations}\label{tylor_equations}
\begin{align}
&\log(\ell_{m}+\sigma^2)\ge\log(\ell_{m,0}+\sigma^2)+\frac{\ell_{m}-\ell_{m,0}}{\ell_{m,0}+\sigma^2},\\
&\ell_{m}\ge \sum_{m'\neq m,m'\in\mathcal{M}}\sum_{d\in\mathcal{K}^{m'}}\rho_{m',d}p_G\text{Tr}({\bf Q}_{m,d} {\bf\Phi}).
\end{align}
\end{subequations}

By substituting~\eqref{intro_var} into the constraint~\eqref{threshold}, it is transformed into a convex form in terms of ${\bf\Phi}$ as follows:
\begin{align}\label{trans_threshold}
&\rho_{m,k}p_G\text{Tr}({\bf Q}_{m,k} {\bf\Phi})
\ge\gamma\Big( \sum\limits_{d>s_m(k),d\in\mathcal{K}^m}\rho_{m,d}p_G\text{Tr}({\bf Q}_{m,d} {\bf\Phi})\notag\\
+&\sum\limits_{m'\neq m,m'\in\mathcal{M}}\sum\limits_{d\in\mathcal{K}^{m'}}\rho_{m',d}p_G\text{Tr}({\bf Q}_{m,d} {\bf\Phi})+\sigma^2\Big).
\end{align}
To this end, we drop the rank-one constraint and adapt passive beamforming matrix ${\bf\Phi}$ by solving the following subproblem:
\begin{equation}\label{question_passive}
\max_{{\bf\Phi}}\sum_{m\in\mathcal{M}}R_m^l~~
\mathrm {s.t.}~~\eqref{equ_channel},\eqref{optimistic_channel}-\eqref{channle_order},\text{ and }\eqref{second_term}-\eqref{trans_threshold}.
\end{equation}
Problem~\eqref{question_passive} is a standard SDR problem and thus is easily tackled by the standard SDR solver. The rank-one solution can be extracted by using the Gaussian randomization method~\cite{Zhao-2023}.

Given the ARIS's passive beamforming $\bm{\theta}$, problem~\eqref{sub_question} becomes an integer programming problem with respect to the GUs' association strategies $\bm{\rho}$. 
We first relax $\bm{\rho}$ to a continuous variable and then rewrite it into an equivalent form as follows:
\begin{subequations}\label{association_trans}
\begin{align}
&\sum_{m\in\mathcal{M}}\sum_{k\in\mathcal{K}}(\rho_{m,k}-\rho_{m,k}^2)\leq 0,   \label{011}    \\
&\sum_{m\in\mathcal{M}}\!\!\rho_{m,k}\leq1 \text{ and } 0 \leq \rho_{m,k} \!\leq\! 1, \forall m\in\mathcal{M},\forall k\in\mathcal{K}.\label{012}
\end{align}
\end{subequations}

We further adopt the penalty method to iteratively approximate the constraint~\eqref{011}. Introducing a penalty coefficient $\kappa$ and an auxiliary variable $\lambda$, we can simply optimize $\bm{\rho}$ by solving the following subproblem:
\begin{subequations}\label{pro_association}
\begin{align}
\max_{\bm{\rho},\lambda}~~&\sum_{m\in\mathcal{M}}R^l_m - \kappa\lambda \\
\mathrm {s.t.}~~&\eqref{equ_channel}, \eqref{optimistic_channel}-\eqref{channle_order}, \eqref{second_term}-\eqref{trans_threshold},\text{ and } \eqref{012}, \\
&\sum_{m\in\mathcal{M}}\sum_{k\in\mathcal{K}}(\rho_{m,k}-\rho_{m,k}^2)\leq \lambda.\label{ori_rho}
\end{align}
\end{subequations}
Constraint~\eqref{ori_rho} can be approximated to a linear form by the first-order Taylor expansion at given point $\rho_{m,k,0}$ as follows:
\begin{equation}
\sum_{m\in\mathcal{M}}\sum_{k\in\mathcal{K}}\rho_{m,k}+ \rho_{m,k,0}^2-2\rho_{m,k}\rho_{m,k,0}\leq \lambda. \label{access_tylor}
\end{equation}
By replacing constraint~\eqref{ori_rho} with the linear approximation~\eqref{access_tylor}, problem~\eqref{pro_association} becomes a standard convex problem.
Problems~\eqref{question_passive} and \eqref{pro_association} are alternately optimized until a stable solution is reached.

The details of the O-HDRL algorithm are summarized in Algorithm~\ref{alg-omaddpg}. Given the ARIS's passive beamforming and the GUs' association strategies, the UAVs' trajectories are updated by using the MADDPG algorithm. Then, following the trajectories, the optimization methods adapt the remaining decision variables in each time slot. To ensure the convergence of the AO algorithm, we compare the solutions of each subproblem in successive iterations and retain the solution with a higher throughput performance. Meanwhile, the O-HDRL framework significantly reduces the action space for all agents, which allows them to search for optimal actions within a lower-dimensional space. This improves not only the stability of convergence but also the overall learning performance.
The complexity of the O-HDRL in each training interaction is characterized as $C_m+C_o$, where $C_m$ and $C_o$ denote the complexities from the MADDPG and optimization modules, respectively. Let $n_{a,f}$ and $n_{c,f}$ represent the number of neurons in the $f$-th layer of the actor and critic networks, respectively. The MADDPG complexity is calculated by $C_m = \mathcal{O}\{\sum_{f=0}^{F_a-1} n_{a,f}n_{a,f+1}+\sum_{f=0}^{F_c-1}n_{c,f}n_{c,f+1}\}$~\cite{Guo-ton2023}, where $F_a$ and $F_c$ denote the layer number of the actor and critic networks, respectively. Meanwhile, the optimization complexity is evaluated by $C_o = \mathcal{O}\{(L^{3.5}+(KM)^{3.5})I_{\max}\}$~\cite{Luo-2010yuan}, where $I_{\max}$ denotes the iteration number introduced by the AO method in lines~\ref{Opt1}-\ref{Opt2} of Algorithm~\ref{alg-omaddpg}.

To improve the overall learning efficiency, we can design the approximate optimization methods to reduce the computation overhead of the optimization part. Note that the optimization solution within the learning process should be computation-efficient but does not have to be very precise. The GUs' association optimization relies on the GUs' channel conditions, which are dynamically controlled by the ARIS's passive beamforming. To decouple their interdependency, we can consider an ideal scenario where the ARIS has the strong capability to align each GU's reflection link with its direct link to the UAV. As such, we can first optimize the GUs' association strategy and then proceed with optimizing the ARIS's passive beamforming strategy. This transforms the AO method into a two-step process, reducing the computational complexity significantly. Furthermore, instead of using the SDR for the ARIS's passive beamforming strategy, we can heuristically align the ARIS's phase shift $\bm{\theta}$ to match the strongest UAV-GU channel within the ARIS's coverage.
Since the GUs with stronger channels are decoded earlier in the SIC process, they always tolerate stronger interference from other GUs, which are decoded later. Thus, aligning the phase shift with the strongest GU-UAV channel makes it easier to meet its decoding requirement.
As such, we only need to optimize the GUs' association strategy $\bm{\rho}$ with the fixed channel conditions, and thus the internal iterations between problems~\eqref{question_passive} and~\eqref{pro_association} can be avoided~\cite{Cui-2023}. The computational complexity of the approximate solution is significantly reduced to $C_o = \mathcal{O}\{L+(KM)^{3.5}\}$.
\begin{algorithm}[t]
	\caption{O-HDRL for UAVs' Trajectory Planning, ARIS's Passive Beamforming, and GUs' Transmission Control}\label{alg-omaddpg}
	\begin{algorithmic}[1]
        \State Initialize the DNN parameters and control variables for all UAVs agents. Set the optimization threshold $\eta\ge0$
        \Statex \textbf{\% Learning for the UAVs' trajectory planning}
        \State \hspace{3mm} Identify the system state ${\bf s}_{i,t}\in\mathcal{S}$
        \State \hspace{3mm} Agents adapt trajectory planning strategies ${\bf q}_m$ and ${\bf q}_r$
        \Statex \textbf{\% Optimizing  ARIS's passive beamforming and}
        \State \hspace{4mm} \textbf{GUs' association}
        \State \hspace{3mm} \textbf{repeat}
        \State \hspace{6mm} Optimize $\bm{\theta}$ by solving subproblem~\eqref{question_passive}\label{Opt1}
        \State \hspace{6mm} Optimize $\bm{\rho}$ by solving subproblem~\eqref{pro_association}\label{Opt2}
        \State \hspace{3mm} \textbf{until} \{$\bm{\theta},\rho_{m,k}$\} reach convergence
        \For{each UAV agent $i \in \mathcal{I}$}
        \State Execute the combined action $\mathbf{a}_{c,i}=\{\mathbf{a}_{l,i},\mathbf{a}_{l}\}$
        \State Observe the reward function $R(\mathbf{s}_i, \mathbf{a}_{c,i})$
        \State Record the transition to the next state ${\bf s}'_{i}$
        \State Store the transition sample $({\bf s}_{i},{\bf a}_{c,i}, R(\mathbf{s}_i, \mathbf{a}_{c,i}), {\bf s}'_{i})$
        \State Update the parameters of actor- and critic-networks
        \EndFor
	\end{algorithmic}
\end{algorithm}
\vspace{-0.3cm}
\section{Dual-mode UAVs with Adaptive Mode Switching}\label{sec-extension}
\begin{figure}[t]
	\centering
	\includegraphics[width = 0.4\textwidth]{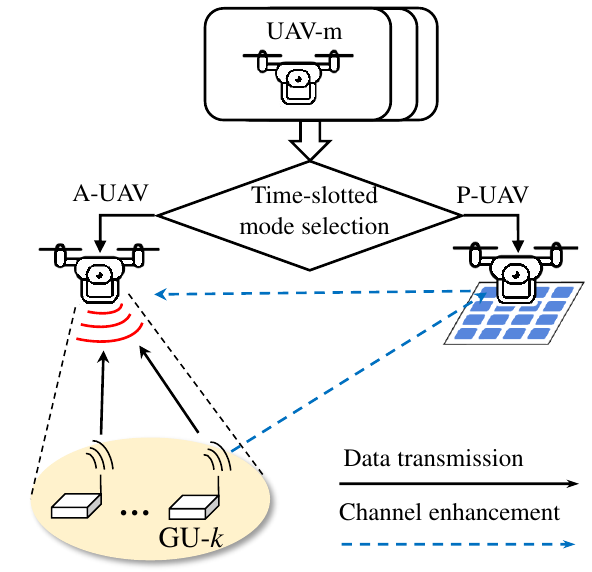}
	\caption{Dual-mode UAV-assisted wireless networks.}\label{fig-struc}
\vspace{-0.4cm}
\end{figure}
Deploying an ARIS can improve the UAVs' channel conditions on its trajectory to provide dynamic services for the GUs' data transmissions. 
Besides the fixed-mode ARIS, we extend it to the dual-mode UAVs to further improve the network flexibility and network capacity. In the dual-mode UAV system, each UAV is equipped with both a passive ARIS and an active antenna system. As such, all UAVs are capable of either passively reflecting (referred to as P-UAV) or actively collecting data (referred to as A-UAV) according to channel conditions and GUs' traffic demands, as shown in Fig.~\ref{fig-struc}. It is expected that the dual-mode UAVs can offer increased flexibility for the network's interference management and capacity enhancement. In particular, when multiple A-UAVs simultaneously collect the GUs' data, each of them suffers co-channel interference from the others. The interference will become stronger as the multiple A-UAVs are in close proximity, which leads to a significant reduction of the overall transmission efficiency. However, if some A-UAVs are capable of switching to the P-UAVs, they can provide channel enhancement instead of the harmful interference signals to the A-UAVs. Meanwhile, multiple P-UAVs can generate collaborative beamforming to achieve higher spatial multiplexing gain for the NOMA GUs' transmissions \cite{Zheng-2021}. 
In another aspect, when the GUs' traffic load becomes unbalanced over the service area, the P-UAVs may also switch back to the A-UAVs to help the heavy-loaded A-UAVs. The UAVs' flexible mode switching on demand can effectively enhance the overall network capacity and throughput.
\begin{figure}[t]
	\centering
	\includegraphics[width = 0.45\textwidth]{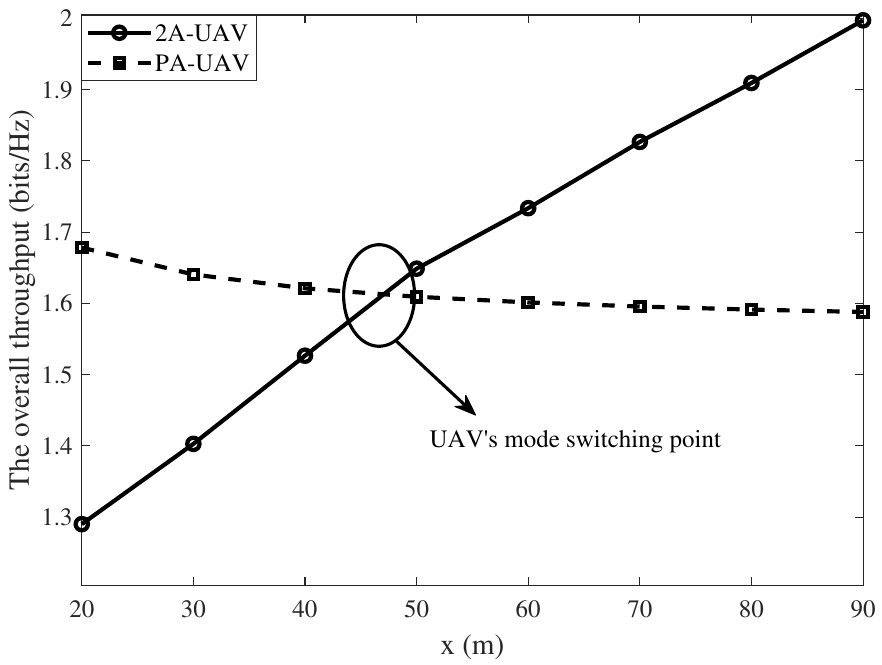}
	\caption{An example for the dual-mode switching scheme.}\label{simulation}
\vspace{-0.6cm}
\end{figure}

To verify the effectiveness of the dual-mode switching scheme, we present an experimental confirmation to compare the overall throughput achieved by two UAVs' operation schemes, i.e, the case with two A-UAVs (denoted by 2A-UAV in Fig.~\ref{simulation}) and the case with a hybrid of P-UAV and A-UAV (denoted by PA-UAV in Fig.~\ref{simulation}). This experiment involves multiple GUs uniformly distributed in the simulation area. We aim to compare the overall throughput achieved by the two-UAV system as we vary the distance between the two UAVs. As depicted in Fig.~\ref{simulation}, as two UAVs are close to each other, the PA-UAV scheme achieves a higher throughput. As the distance increases, the interference between the UAVs decreases, resulting in a higher rate of the 2A-UAV scheme.
It is evident that we can switch the UAVs' operating modes on demand to achieve a better transmission performance. Note that the dual-mode UAVs may introduce additional energy consumption to operate both the active and passive systems. However, we mainly focuses on exploring the transmission improvement by the dual-mode switching scheme, and thus we omit the UAV's energy consumption in this paper. The exploration of the dual-mode UAVs' communication in energy-limited scenarios will be a promising direction for our future research.
\vspace{-0.3cm}
\subsection{Optimizing UAVs' Mode Switching}
We define binary variables $\bm{\alpha}=\{\alpha_m\}_{m\in\mathcal{M}}$ as the UAVs' operation modes, and $\alpha_m=1$ (or $\alpha_m=0$) represents that the UAV-$m$ works at the active (or the passive) mode in the current time slot. Given a mode selection, the equivalent channel from the GU-$k$ to the A-UAV-$m$ is represented as follows:
\begin{equation}
\widehat{{\bf h}}_{m,k}\!=\!\!\!\!\!\!\!\underbrace{\sum_{m'\neq m, m'\in \mathcal{M}}\!\!\!\!\!\!(1-\alpha_{m'}){\bf h}_{m,m'}\text{diag}(\tilde{{\bf h}}_{m',k})\widehat{\bm{\theta}}_{m'}}_{\text{Channel enhancement by P-UAVs}}\!+h_{m,k},\label{mode-switch channel}
\end{equation}
where $\widehat{\bm{\theta}}_{m}$ denote the P-UAV-$m$'s passive beamforming. The first term of the equivalent channel \eqref{mode-switch channel} is enhanced by the P-UAVs' passive beamforming. The UAVs' dynamic mode switching introduces increased controllability to the wireless channel. Hence, it has the potential to improve the channel diversity for the NOMA decoding process and also create more access opportunities for GUs.

Given the mode switching strategy $\bm{\alpha}$, the data rate of the A-UAV-$m$ from the GU-$k$ is represented as follows:
\begin{equation}\label{mode-switch-rate}
\widehat{r}_{m,k} =\alpha_{m}\log(1+\widehat{\text{SINR}}_{m,k}),
\end{equation}
where $\widehat{\text{SINR}}_{m,k}$ is obtained by introducing the equivalent channel~\eqref{mode-switch channel} into~\eqref{SINR-N}.

By jointly optimizing the dual-mode UAVs' trajectory planning ${\bf q}_m$ and mode switching ${\bm \alpha}$, passive beamforming $\widehat{\bm{\theta}}_m$ as well as the GUs' transmission control $\bm{\rho}$ strategies, the throughput maximization problem is updated as follows:
\begin{subequations}\label{reformulated_question}
\begin{align}
\max_{{\bm \alpha},{\bf q}_m,\widehat{\bm{\theta}}_m,\bm{\rho}}&\sum_{n\in\mathcal{N}}\sum_{m\in\mathcal{M}}\sum_{k\in\mathcal{K}}\min\{\tau \widehat{r}_{m,k},D_k\},~~\\
\mathrm {s.t.}~~&\eqref{mobility_safety}, \eqref{association}-\eqref{data_queue},\text{ and }\eqref{mode-switch channel}-\eqref{mode-switch-rate}.
\end{align}
\end{subequations}
The coupling of the mode switching and passive beamforming in~\eqref{mode-switch channel} makes problem~\eqref{reformulated_question} challenging to solve. We first reformulate it introducing the following auxiliary matrices:
\begin{subequations}\label{reformulated_channel}
\begin{align}
&{\bf h}_{m}=[(1-\alpha_{1}){\bf h}_{m,1}, \ldots, (1-\alpha_{m'}){\bf h}_{m,m'}],\\
&\overline{{\bf h}}_{m,k}=[\tilde{{\bf h}}_{1,k},\ldots, \tilde{{\bf h}}_{m',k}],\\
&\tilde{\bm{\theta}}_m=[\widehat{\bm{\theta}}_1,\ldots,\widehat{\bm{\theta}}_{m'}],
\end{align}
\end{subequations}
where $m'\neq m$ and $m,m'\in\mathcal{M}$. As such, the equivalent channel $\widehat{{\bf h}}_{m,k}$ in~\eqref{mode-switch channel} can be reformulated as follows:
\begin{equation}\label{reformulated_equivalent channel}
\widehat{{\bf h}}_{m,k} ={\bf h}_{m}\text{diag}(\overline{{\bf h}}_{m,k})\tilde{\bm{\theta}}_{m} + h_{m,k}.
\end{equation}
It is seen that the equivalent channel $\widehat{{\bf h}}_{m,k}$ in~\eqref{reformulated_equivalent channel} has a similar structure to~\eqref{equ_channel}. Therefore, given the mode switching ${\bm \alpha}$, we can optimize the dual-mode UAVs' trajectories ${\bf q}_m$, the passive beamforming $\tilde{\bm{\theta}}_m$, and the GUs' transmission control $\bm{\rho}$ in problem~\eqref{reformulated_question} by using the similar method as proposed in the Section~\ref{subproblem-opti}. Thus, the main difference between the fixed-mode ARIS and dual-mode UAV systems lies in searching for the optimal dual-mode switching strategy. From~\eqref{mode-switch channel}, we can observe that as more dual-mode UAVs switch to the P-UAVs, the number of the available reflecting links increases, which can potentially enhance the channel conditions for the A-UAVs. However, this will also reduce the number of the A-UAVs for data collection, as indicated in~\eqref{mode-switch-rate}. As such, it is challenging to directly determine the UAVs' mode-switching strategy using a specific closed-form solution.
\vspace{-0.3cm}
\subsection{Learning Mode Switching between P-UAV and A-UAV}
Intuitively, we can list all possible combinations of the UAVs' operating modes and formulate the optimization problem for each of them. Then, each optimization problem can be solved by Algorithm~\ref{alg-omaddpg} proposed in the Section~\ref{subproblem-opti}. However, this becomes computationally infeasible, particularly when dealing with a large number of UAVs. Hence, we also employ the MADDPG to update the mode-switching strategy. To proceed, the learning action $\mathbf{a}_{l,m}$ of each agent is modified to include the dual-mode UAVs' next position ${\bf q}_m$ and the mode-switching strategy $\bm{\alpha}$. The design of the reward function should reflect  the dual-mode UAVs' different objectives when they dynamically select the different operation modes. Hence, the mode-related reward function $R^D_{m}$ is represented as follows:
\begin{equation}\label{reward-dualmode}
R^D_{m} =\alpha_m R_{m,a} + (1-\alpha_m)R_{m,p},
\end{equation}
where $R_{m,a}$ and $R_{m,p}$ denote the rewards of the A-UAV-$m$ and P-UAV-$m$, respectively. The A-UAV-$m$'s reward $R_{m,a}$ aims to optimize the data transmission performance, while the P-UAV-$m$'s reward $R_{m,p}$ is designed to improve the channel conditions for all A-UAVs. The expressions of $R_{m,a}$ and $R_{m,p}$ are similar to those in~\eqref{reward_FIXARIS}.
\begin{figure}[t]
	\centering
	\includegraphics[width = 0.45\textwidth]{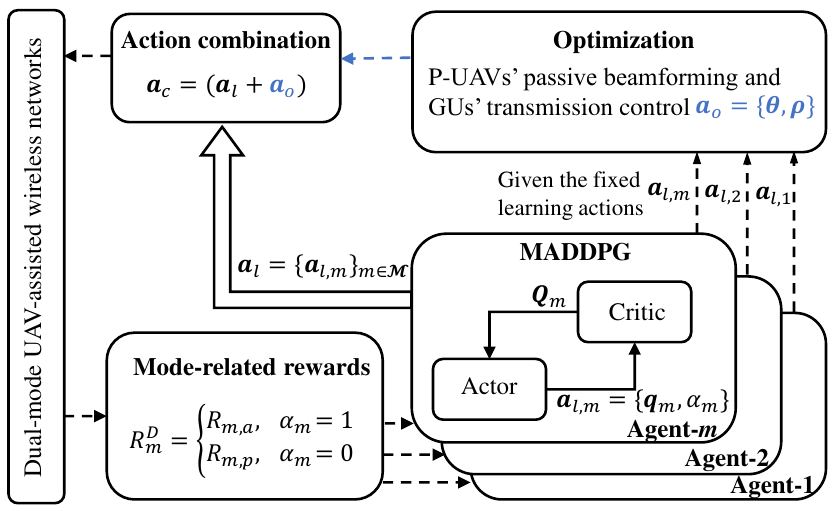}
	\caption{The O-HDRL for dual-mode switching scheme.}\label{diagram2}
\vspace{-0.6cm}
\end{figure}
We design a learning framework, as shown in Fig.~\ref{diagram2}. In each decision slot, the multiple agents first make decisions on the dual-mode UAVs' trajectories and mode switching strategies. Then, each agent adjusts the passive beamforming and the transmission control strategies by the optimization method. By combining the actions and interacting with the environment, each dual-mode UAV can gradually refine its decision and chooses a preferable operation mode.
\vspace{-0.4cm}
\section{Numerical Results}\label{sec-simulation}
\begin{table}[t]
\caption{Parameter settings in the simulations.} \label{para_settings}\normalsize
	\centering
	\begin{tabular}{|l|l|}
		\hline
        Parameters&Settings\\
        \hline
		GU's transmit power $p_G$ & $30$ dBm \\
		Number of the AIRS's elements $L$& $30$\\
		Flying altitude of the UAVs $H$ & $30$ m\\
        Background noise power $\sigma$& $-90$ dBm\\
		Maximum speed of the UAVs $V_{\text{max}}$& $50$ m/s\\
        Safety distance to the UAVs $D_{\text{min}}$ & $5$ m\\
        Number of time slots $N$ & $50$\\
        Actor's learning rate& $10^{-4}$\\
        Critic's learning rate&$10^{-3}$\\
        Reward discount factor &$0.95$\\
        Size of the replay buffer &$5\times10^{5}$\\
        Size of the mini-batch &$256$\\
        $\epsilon$-greedy coefficient & $0.1$\\
        \hline
	\end{tabular}
\vspace{-0.6cm}
\end{table}
In this section, numerical results are presented to evaluate the performance of the proposed ARIS-assisted and dual-mode switching schemes (denoted as Fixed-ARIS and DM-Switching, respectively) for the multi-UAV wireless networks. We consider $M=3$ UAVs collecting data from $K=9$ GUs randomly distributed in a $(1000 \times 1000)$ m$^2$ area, similar to that in~\cite{Zeng-2019twc}. The UAV-GU and the UAV-UAV links follow the Rican and LoS channel models, respectively. Other parameters are set as in  Table \ref{para_settings}, similar to those in~\cite{Hua-2022new}.
\vspace{-0.4cm}
\subsection{Convergence of the O-HDRL Framework}
\begin{figure}[t]
	\centering
	\subfloat[Convergence performance of the optimization in one time slot.]{\includegraphics[width=0.43\textwidth]{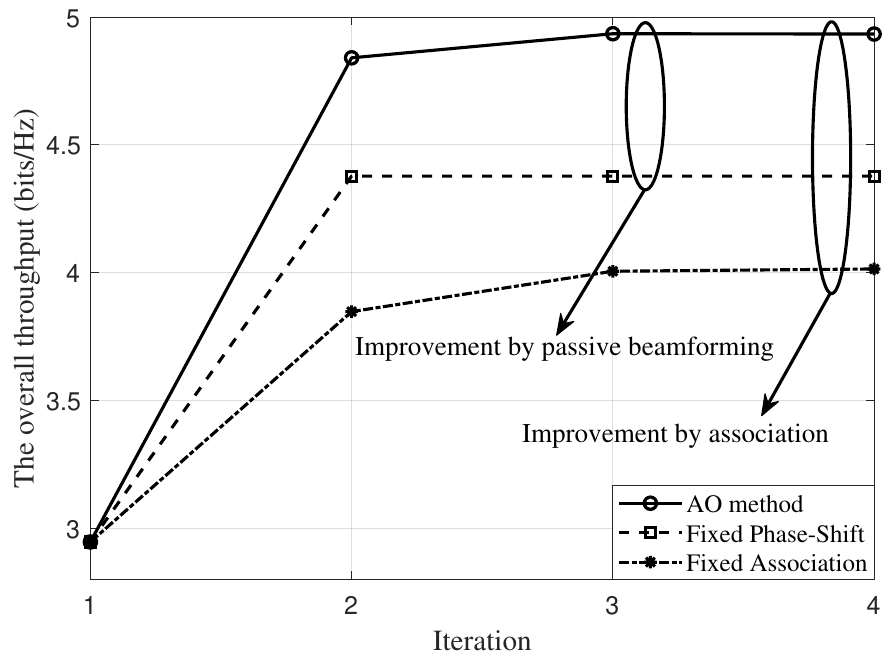}}\label{convergence-optimization}
	\subfloat[Convergence performance of the O-HDRL algorithms.]{\includegraphics[width=0.43\textwidth]{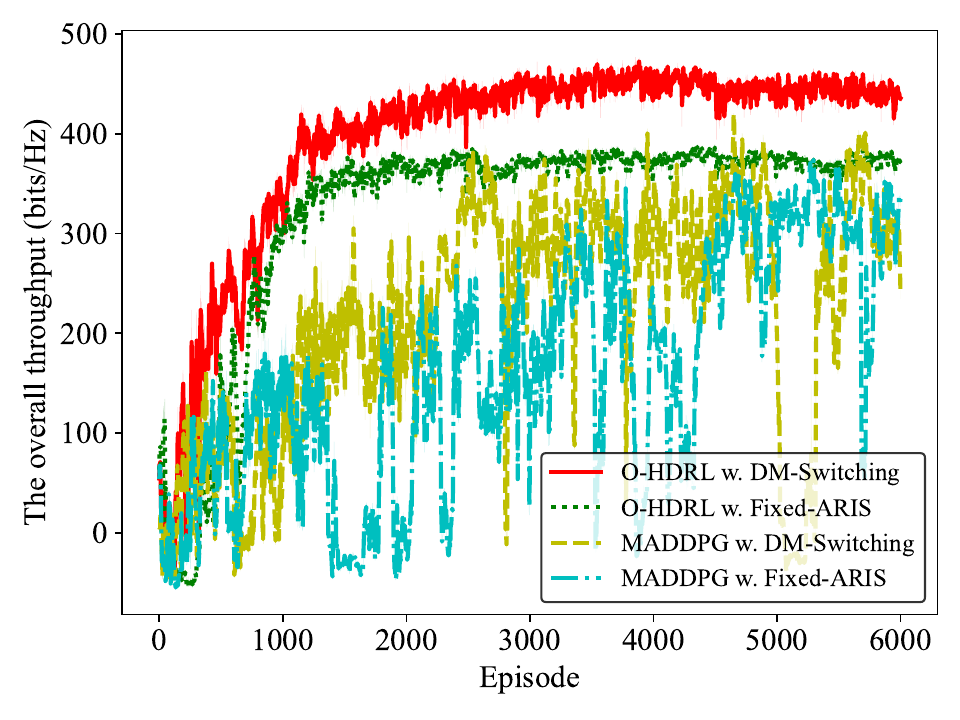}}\label{convergence-overall}
	\caption{Optimization improves the learning performance.}\label{fig-convergence}
\vspace{-0.7cm}
\end{figure}
In Fig.~\ref{fig-convergence}, we validate the convergence performance of the proposed O-HDRL framework. In the multi-agent learning process, the optimization is continually introduced to refine the policies of the agents. In each time slot, the GUs' association and ARISs' passive beamforming strategies are jointly optimized by using the AO method. Fig.~\ref{fig-convergence}(a) illustrates the achievable throughput of different optimization schemes as the algorithm iterates. The AO method is compared with two benchmarks: (\romannumeral 1) Fixed Phase-Shift scheme where only GUs' association strategy is optimized with the fixed ARISs' passive beamforming, and (\romannumeral 2) Fixed Association scheme where only the ARISs' passive beamforming strategy is optimized while keeping the GUs' association strategy fixed. The AO method achieves a higher throughput compared to the benchmarks. This validates that the joint optimization of the GUs' association and the ARISs' passive beamforming strategies improves the data transmission efficiency. Furthermore, the AO method shows a rapid convergence, which verifies that subproblems~\eqref{question_passive} and \eqref{pro_association} are efficiently tackled.

Figure~\ref{fig-convergence}(b) reveals the convergence and reward performance of the O-HDRL framework with different UAVs' operation modes compared to the conventional MADDPG. In the MADDPG, all control variables are learned by the DNN at each decision-making slot.
It is observed that the O-HDRL algorithms achieve superior stability and convergence performance compared to the MADDPG algorithm. Since the MADDPG needs to simultaneously explore the ARISs' passive beamforming and the GUs' association strategies, it has a much higher dimension for the action space compared to that of the O-HDRL. The high-dimensional action space introduces instability to the multi-agent's learning process, which makes it more challenging to explore the optimal policy. For the O-HDRL algorithms, the action space is effectively reduced by introducing the optimization modules.
Furthermore, by using the O-HDRL framework, the DM-Switching scheme provides $17.51\%$ higher throughput than that of the Fixed-ARIS scheme. This validates that the UAVs' mode switching on demand can better adapt to dynamic wireless networks.

\begin{figure}[t]
	\centering
	\subfloat[Fixed-ARIS scheme.]{\includegraphics[width=0.22\textwidth]{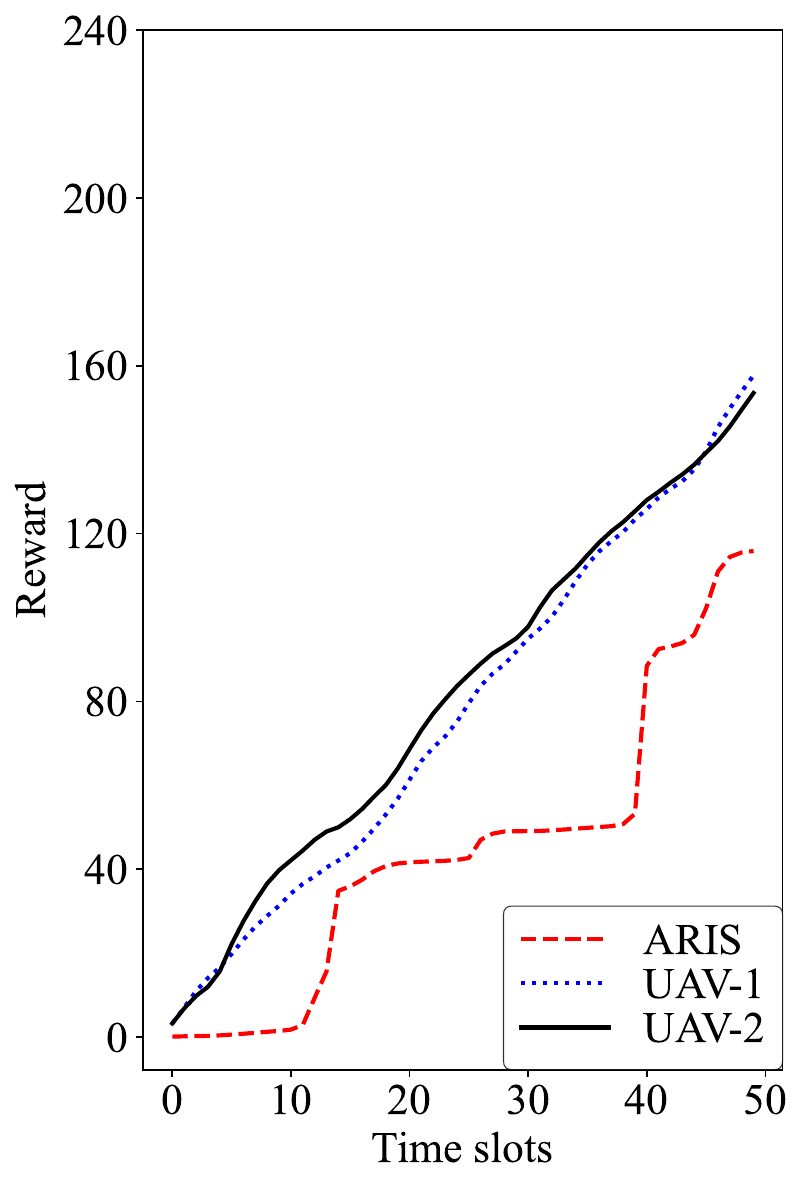}}\label{aris-reward}
	\subfloat[DM-Switching scheme.]{\includegraphics[width=0.22\textwidth]{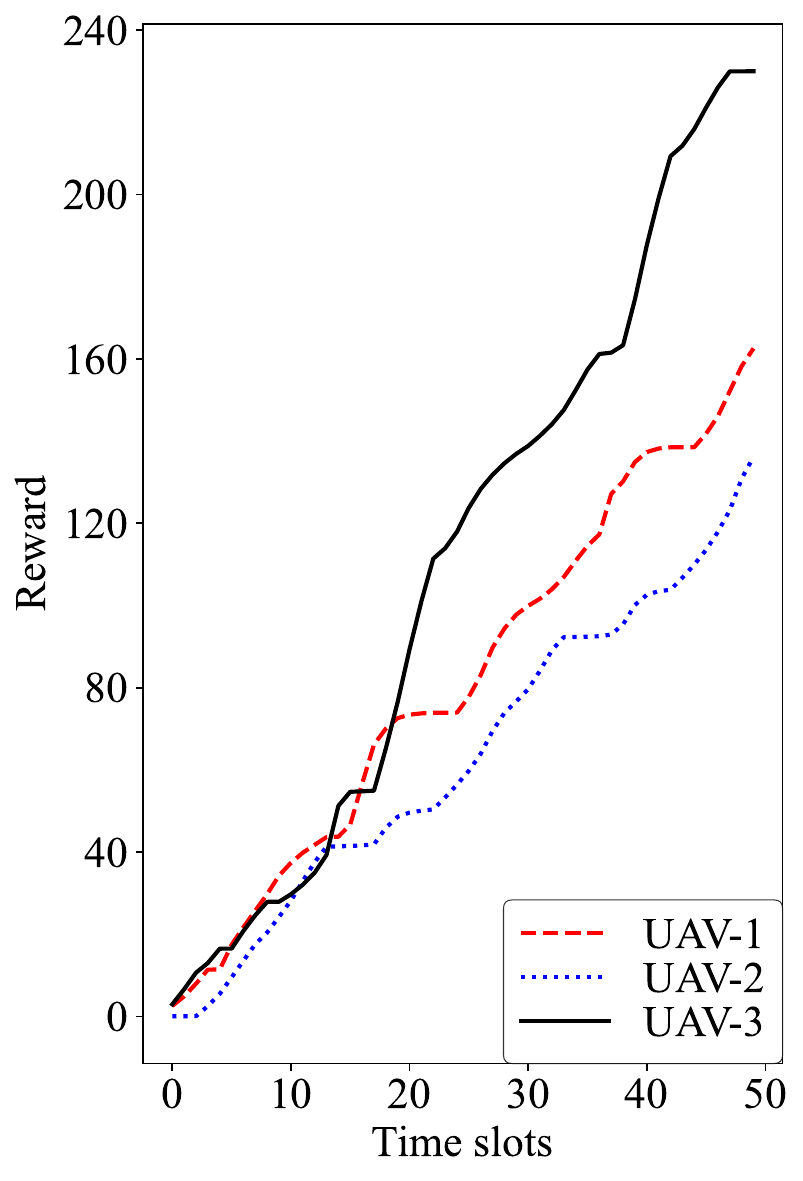}}\label{dmuav-reward}
	\caption{Agents' reward improvement in different time slots. }\label{each-reward}
\vspace{-0.5cm}
\end{figure}

Figure~\ref{each-reward} shows the agents' reward performance of the Fixed-ARIS and the DM-Switching schemes. The Fixed-ARIS scheme's reward performance is shown in Fig.~\ref{each-reward}(a). It is observed that the ARIS's reward increases in a stepwise manner. This is because the ARIS needs to dynamically fly to a new position to provide channel reflection for the UAVs. When the UAVs change their service places, the ARIS should also move accordingly to achieve considerable channel enhancement. Instead, the DM-Switching scheme can switch UAVs' operating modes according to the channel conditions. When P-UAVs cannot obtain good rewards by providing the channel enhancement for the other A-UAVs, they can switch to A-UAVs to continue improving their rewards by collecting data. As such, the DM-Switching scheme can make more flexible decisions and thus achieve continuous growth in reward performance, as shown in Fig.~\ref{each-reward}(b).
\vspace{-0.4cm}
\subsection{Trajectory Planning with Fixed-ARIS and DM-Switching}
\begin{figure}[t]
	\centering
	\subfloat[Uniform GUs' distribution.]{\includegraphics[width=0.24\textwidth]{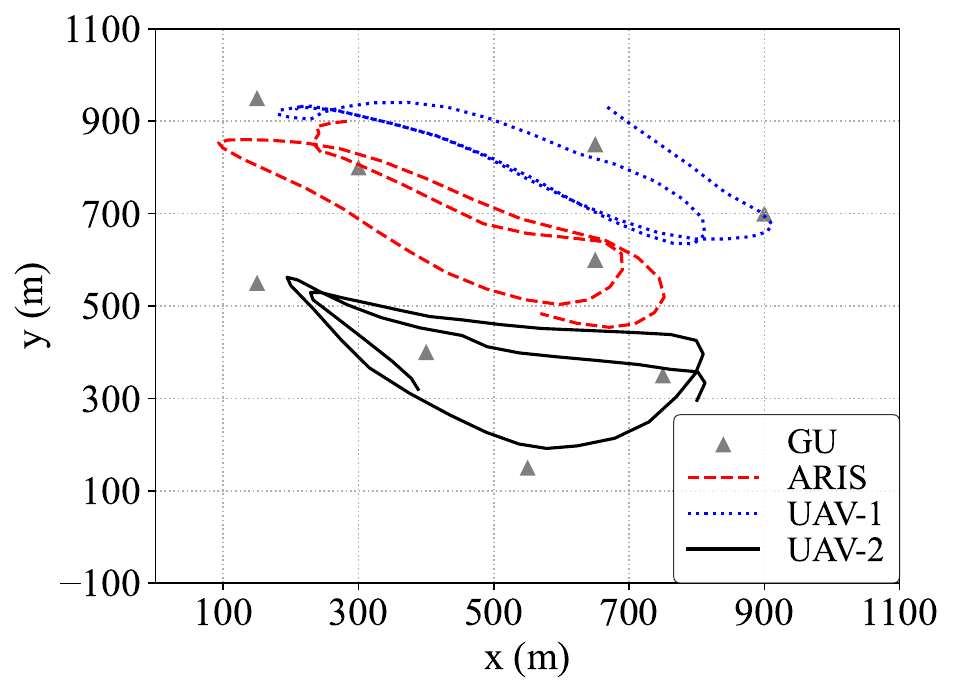}}\label{tra-uniform}
	\subfloat[Non-uniform GUs' distribution.]{\includegraphics[width=0.24\textwidth]{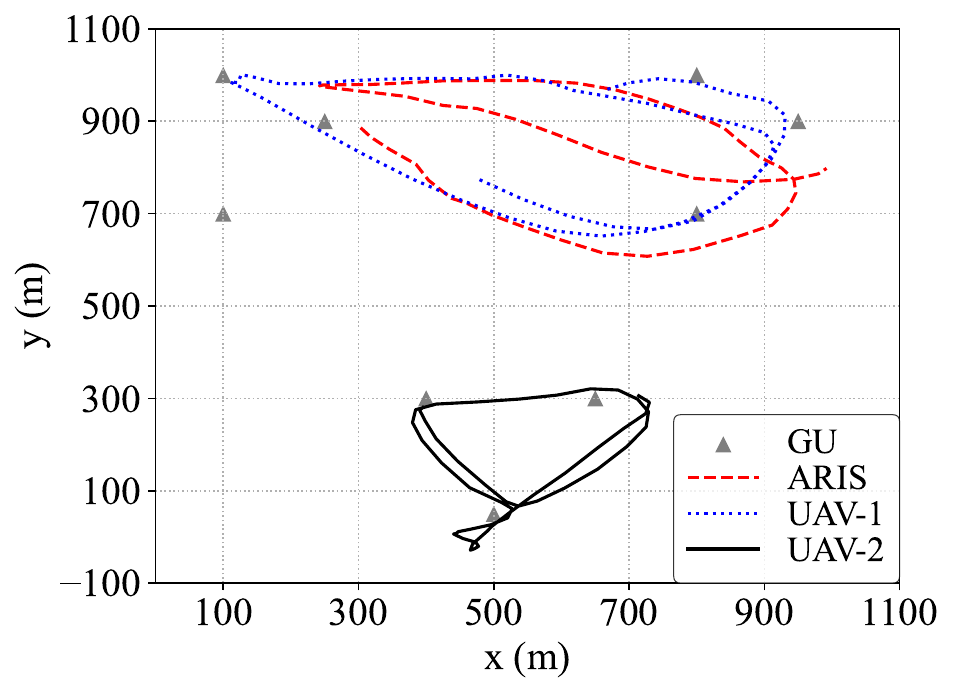}}\label{tra-clustered}
	\subfloat[Uniform GUs' distribution.]{\includegraphics[width=0.24\textwidth]{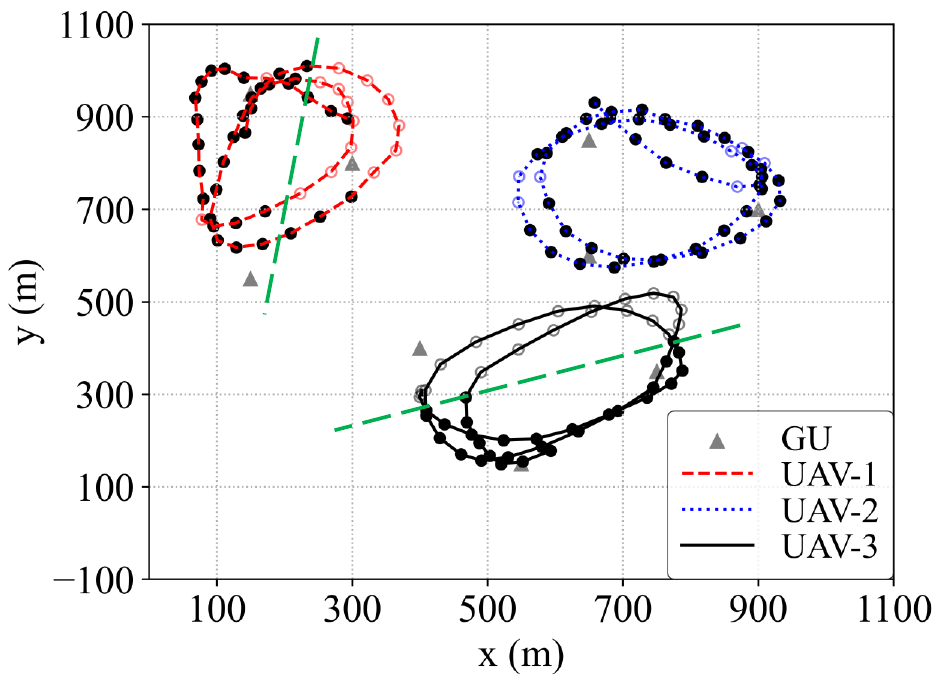}}\label{tra-uniform-uav}
	\subfloat[Non-uniform GUs' distribution.]{\includegraphics[width=0.24\textwidth]{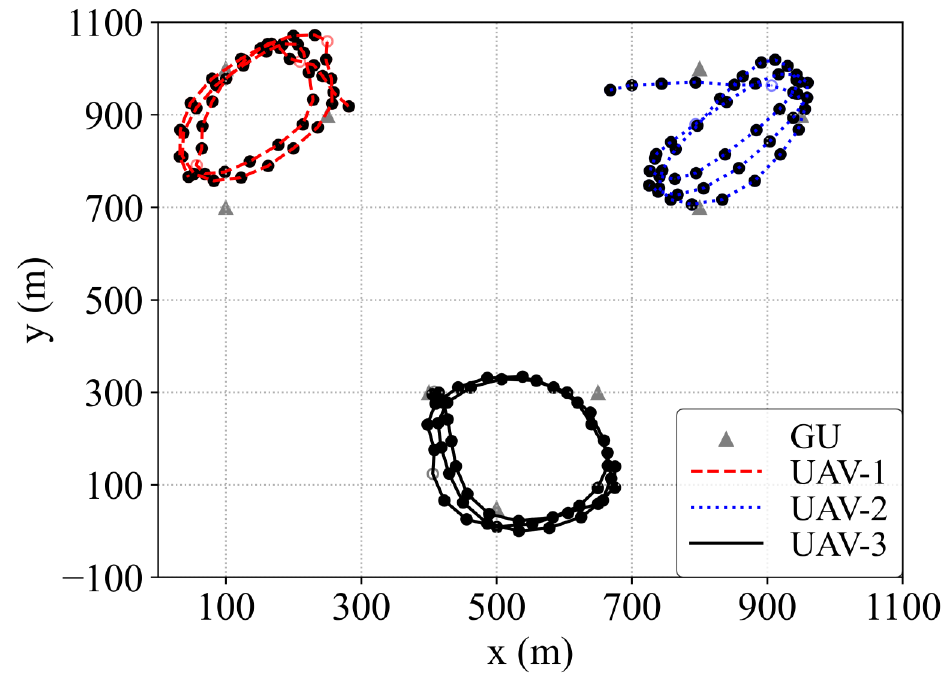}}\label{tra-clustered-uav}
\subfloat[Fixed-ARIS scheme.]{\includegraphics[width=0.24\textwidth]{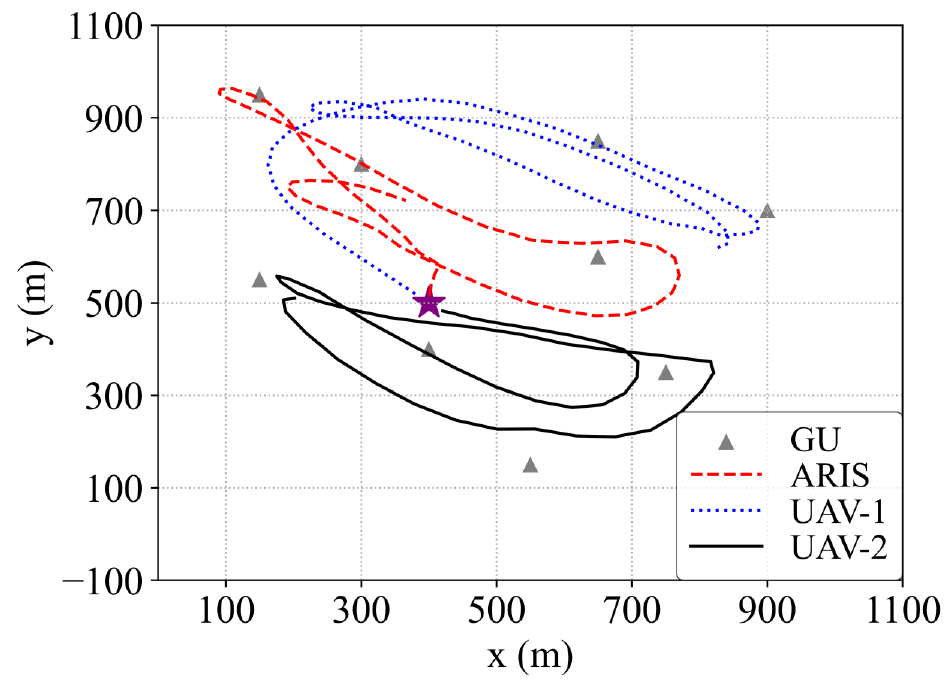}}\label{tra-sameposaris}
    \subfloat[DM-Switching scheme.]{\includegraphics[width=0.24\textwidth]{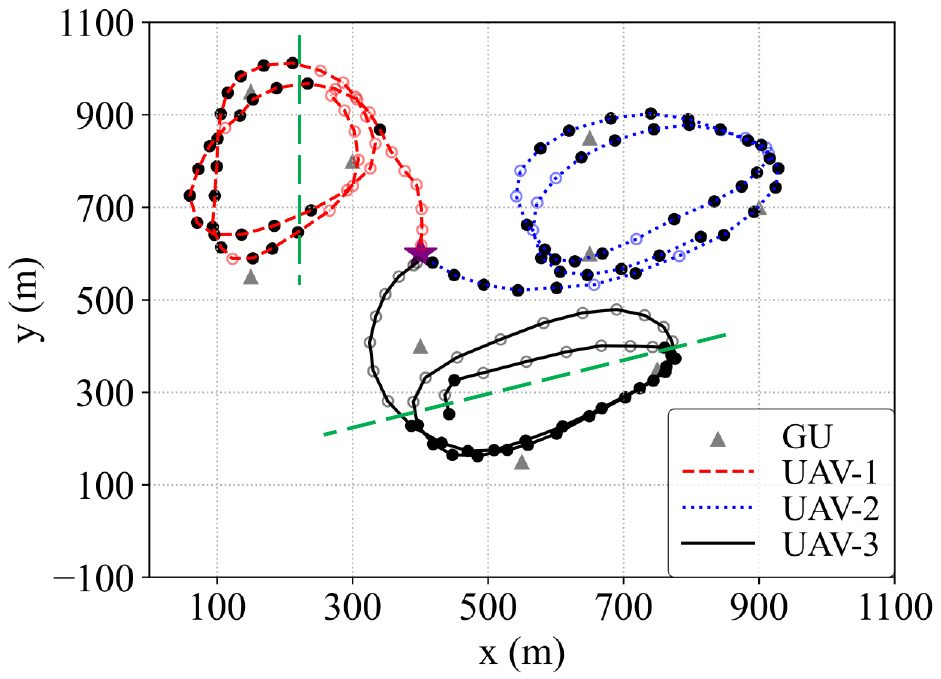}}\label{tra-sameposuav}
	\caption{UAVs' trajectories in Fixed-ARIS and DM-Switching.}\label{fig-trajectories}
\vspace{-0.7cm}
\end{figure}

We present the UAVs' trajectories of the Fixed-ARIS and DM-Switching schemes.
We study the UAVs' trajectory planning in two different spatial distributions of the GUs, i.e., the uniform and non-uniform distributions of the GUs. In the non-uniform case, there are a few points of interests (PoIs), which are the cluster centers for the GUs' spatial distribution. The GUs' locations are generated using the Gaussian distribution centered at these PoIs as shown in Fig.~\ref{fig-trajectories}(b) and (d).

Figure~\ref{fig-trajectories}(a) shows the trajectory planning for the Fixed-ARIS scheme under the uniform GUs' distribution. The UAVs and the ARIS individually adapt their flying locations based on the current observation of the environment. The trajectory of the ARIS is denoted in red dashed line and the trajectories of the UAVs are marked in blue and black solid lines, respectively. The UAVs can autonomously categorize the GUs into two distinct serving groups according to the GUs' locations. Such a spatial division of the data transmission allows each UAV to focus on the GUs within a designated region, which can reduce the co-channel interference from the other UAVs. An interesting observation is that the ARIS flies between the trajectories of the two UAVs, which provides the channel enhancement and/or interference mitigation for both UAVs simultaneously.
Fig.~\ref{fig-trajectories}(b) reveals the trajectory planning under the non-uniform distribution of the GUs. The UAVs also divide the GUs into two groups, and each UAV is responsible for collecting data from its own region. Interestingly, the trajectory of the ARIS is always close to the UAV-$1$ assisting its data transmissions. This is because when two UAVs are far from each other, it is challenging for the ARIS to simultaneously enhance the channels of both UAVs. As such, the ARIS tends to assist the UAV-$1$, which has a higher data transmission load.

Figures~\ref{fig-trajectories}(c) and (d) illustrate the trajectory planning for the DM-Switching scheme under different GUs' distributions. The dynamic changes of the UAVs' mode switching are indicated by distinct markers on the UAVs' trajectories. The solid and hollow markers represent the A-UAV and the P-UAV, respectively. In the uniform case as shown in Fig.~\ref{fig-trajectories}(c), the dual-mode UAVs also divide the GUs into different service groups similar to that in the Fixed-ARIS scheme. The dual-mode UAVs continuously switch modes along with their trajectories. When the dual-mode UAVs are in close proximity, they tend to operate as the P-UAVs. This behavior is driven by the co-channel interference experienced by the A-UAVs when they are in close proximity. By switching some A-UAVs to the P-UAVs, the interference among the A-UAVs is effectively reduced. Meanwhile, the P-UAVs also improve the channel conditions for the A-UAVs.
On the contrary, as the dual-mode UAVs fly apart, they will switch back to the A-UAVs to collect data individually from the GUs. We use green dashed lines to broadly divide the mode switching between the A-UAVs and the P-UAVs. It is evident that the UAVs dynamically switch their modes based on the distances between them and the distribution of the GUs' traffic demands. These results verify that the UAVs' dual-mode switching scheme can adaptively handle the dynamic network environment.

Figure~\ref{fig-trajectories}(d) shows the dual-mode UAVs' trajectory planning in the non-uniform case. The dual-mode UAVs mainly operate as the A-UAVs for data transmissions during the overall trajectories. This is attributed to the large distances between the UAVs, which have reduced interference to each other. Thus, there is no need for the UAVs to switch to the P-UAVs for mitigating the interference. The dual-mode UAVs prefer to operate as the A-UAVs to independently collect the data from the GUs. We further validate the trajectory planning by the Fixed-ARIS and the DM-Switching schemes starting from the same initial location, as shown in Fig.~\ref{fig-trajectories}(e) and (f). The same initial state location implies that UAVs need greater autonomous capability to divide the GUs and plan reasonable trajectories to ensure the fair services for all GUs. We use purple star to represent the UAVs' initial locations. The UAVs in the Fixed-ARIS and DM-Switching scheme can autonomously serve different GU groups and adjust their operation modes according to GUs' spatial locations and channel conditions. This validates that the O-HDRL can balance the UAVs' resource allocation across different regions.
\vspace{-0.5cm}
\subsection{Throughput with the Fixed-ARIS and DM-Switching}
\begin{figure}[t]
	\centering
	\includegraphics[width = 0.45\textwidth]{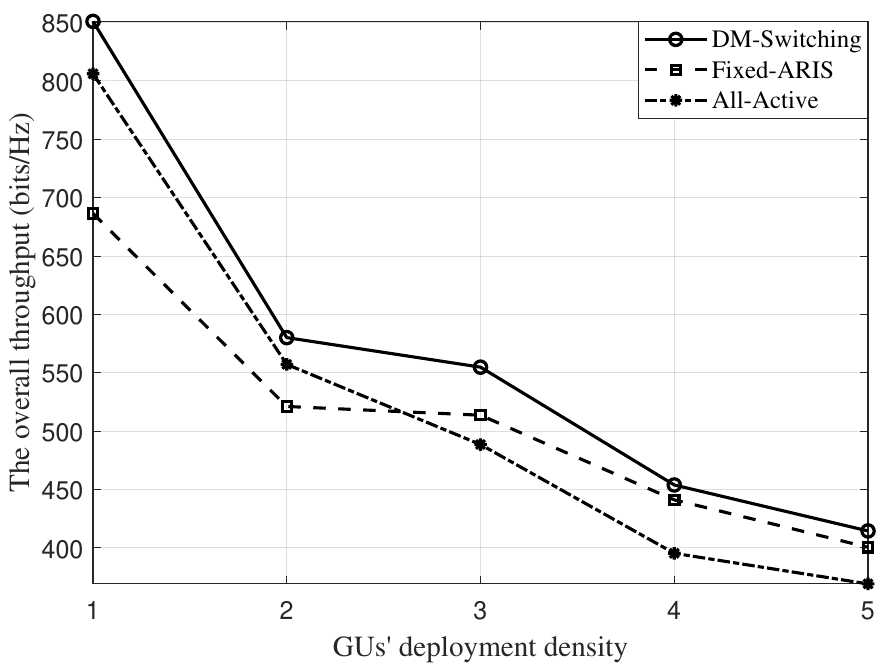}
	\caption{The impact of GUs' deployment on the throughput.}\label{density}
\vspace{-0.5cm}
\end{figure}
We delve into the throughput improvement by comparing the DM-Switching, Fixed-ARIS, and All-Active schemes under different system parameters, where the All-Active scheme represents that all dual-mode UAVs keep working as A-UAVs. 
We investigate the impact of different interference levels on throughput performance by varying GUs' deployment density. We first normalize the GUs' deployment density and increase the density from $1$ to $5$. 
As shown in Fig.~\ref{density}, the overall throughput of all schemes decreases with the increase of the GUs' density. This is because the increased interference diminishes the NOMA transmission efficiency and enforces the competition for GUs' access. When the GUs' deployment density is lower than $2$, the All-Active scheme exhibits a higher throughput than the Fixed-ARIS scheme. As the GUs' density increases, the throughput performance of the Fixed-ARIS scheme surpasses that of the All-Active scheme. This is because the All-Active scheme efficiently utilizes A-UAVs to collect data from the GUs under low interference and good channel conditions.
However, as GUs' density increases, the interference becomes more severe, which intensifies channel competition. In contrast, the Fixed-ARIS scheme employs the ARIS to refine the channel conditions to lower the interference, achieving a higher throughput in a high-density scenario.
Importantly, the DM-Switching scheme achieves the highest overall throughput due to the mode switching on demand. In particular, when the GUs' deployment density is $1$, the DM-Switching scheme improves the throughput by $23.97\%$ compared to the Fixed-ARIS scheme. When the GUs' deployment density is $5$, the throughput enhancement of the DM-Switching scheme is $12.29\%$ compared to the All-Active scheme. 

\begin{figure}[t]
	\centering
	\includegraphics[width = 0.45\textwidth]{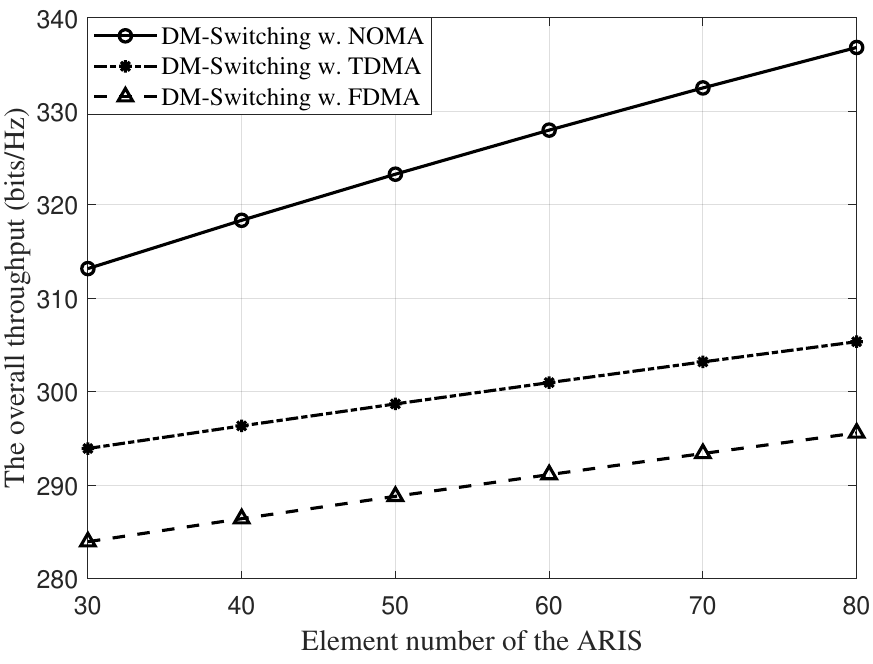}
    \caption{The impact of the ARIS's size on the throughput.}\label{Size}
    \vspace{-0.5cm}
\end{figure}
Different access methods have unique characteristics and affect throughput performance in different ways. Besides NOMA method, other effective access methods, such as time division multiple access (TDMA) and frequency division multiple access (FDMA), can separate GUs in time and frequency domains, respectively.
To explore which access method is more suitable, we focus on the DM-Switching scheme under different GUs' access methods, i.e., NOMA, TDMA, and FDMA, which are denoted as DM-Switching w.~NOMA, DM-Switching w.~TDMA, and DM-Switching w.~FDMA, respectively. In Fig.~\ref{Size}, we investigate the overall throughput as the number of the ARISs' elements $L$ varies from $30$ to $80$. It is observed that all schemes' throughput performance increases with the increased ARISs' elements. This is because the large-size ARISs enhance channel conditions efficiently. The NOMA method exhibits superior performance compared to TDMA and FDMA methods. This is because the additional channel orthogonality introduced by the large-size ARISs  allowing the NOMA GUs to better utilize the channel resources.

\begin{figure}[t]
	\centering
	\includegraphics[width = 0.45\textwidth]{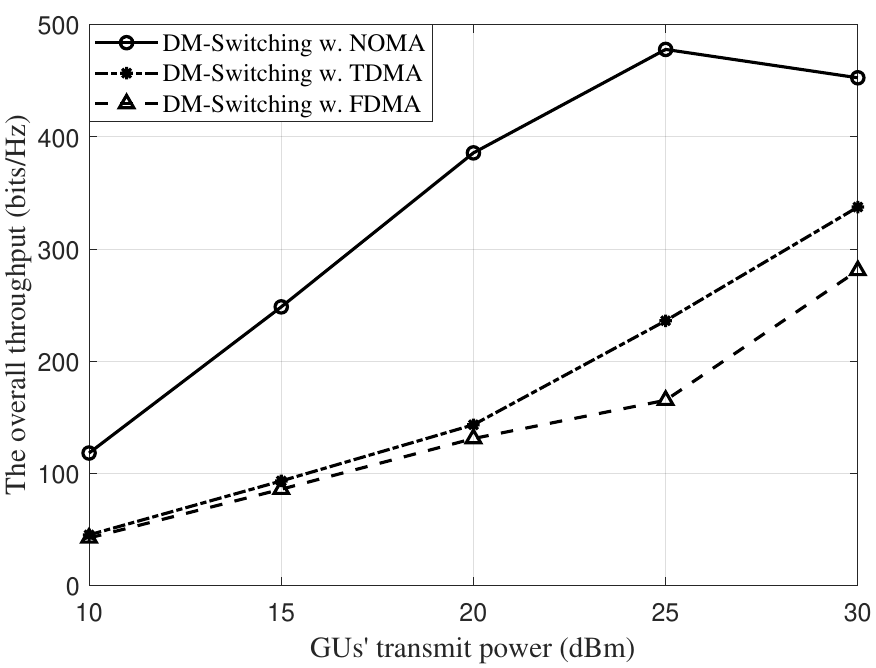}
	\caption{The impact of GUs' transmit power on the throughput.}\label{power}
\vspace{-0.5cm}
\end{figure}
In Fig.~\ref{power}, we further investigate the throughput performance of the DM-Switching scheme under different GUs' transmit power. As GUs' transmit power increases, the interference among GUs becomes more severe. It is shown that the throughput performance of the NOMA scheme surpasses that of the TDMA and the FDMA schemes.
As the GUs' transmit power continuously increases, the throughput performance of the NOMA scheme initially rises and then reduces. This is because given a low transmit power of the GUs, the NOMA method allows multiple GUs to transmit data simultaneously without much mutual interference. However, as the interference among GUs becomes stronger, it becomes hard to meet the SIC decoding requirements, resulting in a decline in the overall throughput performance.
\vspace{-0.2cm}
\section{Conclusions}\label{sec-conclusion}
\vspace{-0.1cm}
In this paper, we have investigated a multi-UAV-assisted NOMA transmission system. A fixed-mode ARIS scheme has been proposed to reduce the co-channel interference among the UAVs and improve the channel diversity for the GUs' NOMA transmissions. The ARIS brings greater flexibility into the channel adjustments, enabling the multi-UAV systems to better adapt to dynamic network environments. We have further proposed a dual-mode switching scheme to better deal with network dynamics. The dual-mode UAVs adaptively switch between the A-UAV mode and the P-UAV mode for either data transmissions or channel improvement according to the network dynamics. We have maximized the overall throughput in the proposed two schemes by jointly optimizing the UAVs' trajectory planning and mode switching, the ARIS's passive beamforming, and the GUs' transmission control strategies. We have proposed an O-HDRL framework that integrates the advantages of the conventional model-free DRL and the optimization methods. Simulation results have demonstrated that the O-HDRL algorithm significantly improves the learning stability and performance compared to the benchmark methods. Moreover, the dual-mode switching scheme has shown a superior throughput performance compared to the fixed-mode ARIS scheme, especially in high-density networks.
\vspace{-0.3cm}

\footnotesize

\bibliographystyle{IEEEtran}
\bibliography{dual-mode-uav}
\begin{IEEEbiography}[{\includegraphics[width=1in,height=1.25in,clip,keepaspectratio]{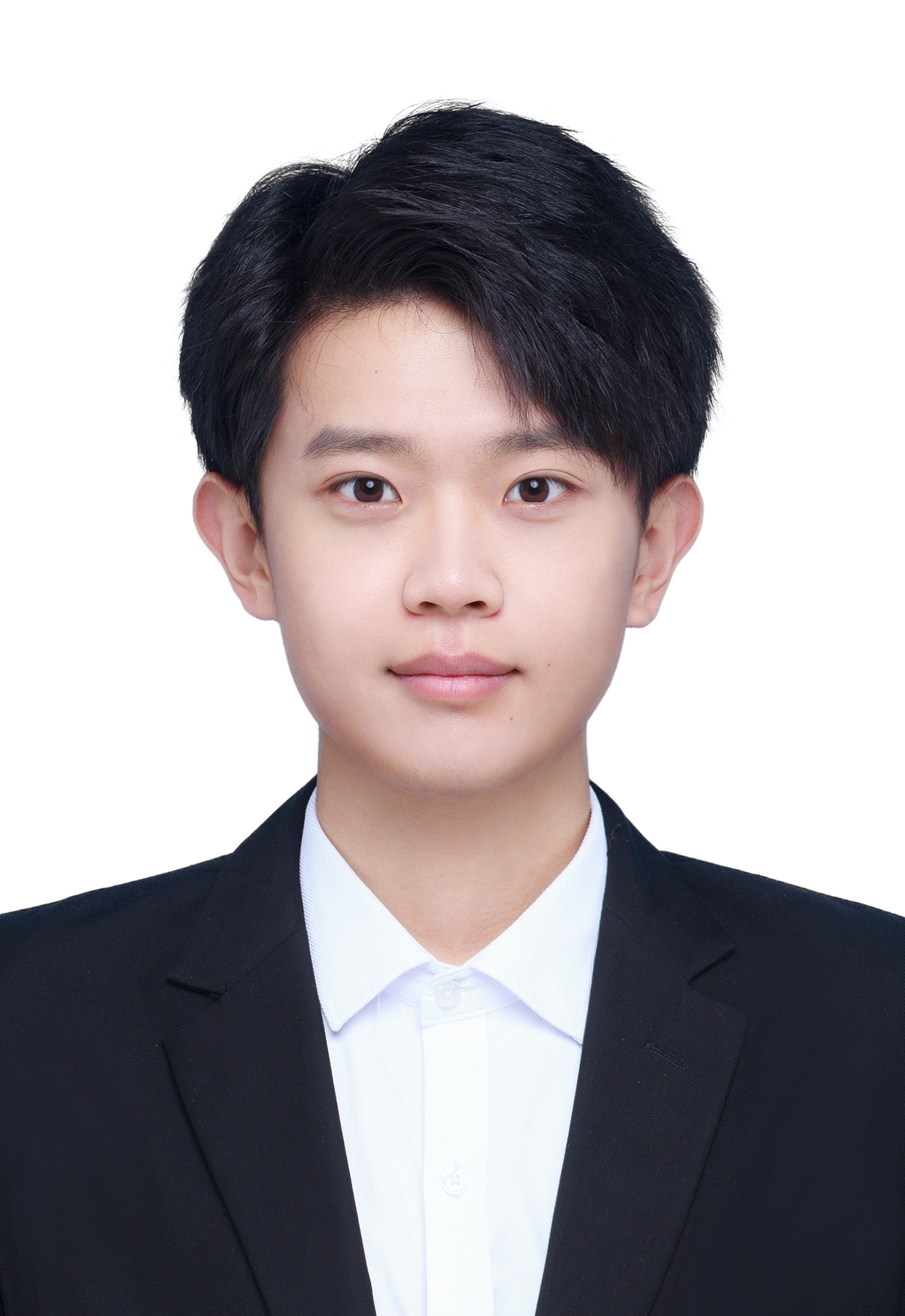}}]{Songhan Zhao} received the B.E. degree in automation from Hefei University of Technology, Heifei, China, in 2018, and the M.E. degree in control engineering from Yanshan University, Qinghuangdao, China, in 2022. He is currently pursuing the Ph.D. degree with the School of Intelligent Systems Engineering, Sun Yat-sen University, Shenzhen, China. His research interests include optimization methods, machine learning, semantic communications, and IRS/UAV-assisted wireless networks.
\vspace{-0.6cm} 	
\end{IEEEbiography}
\begin{IEEEbiography}[{\includegraphics[width=1in,height=1.25in,clip,keepaspectratio]{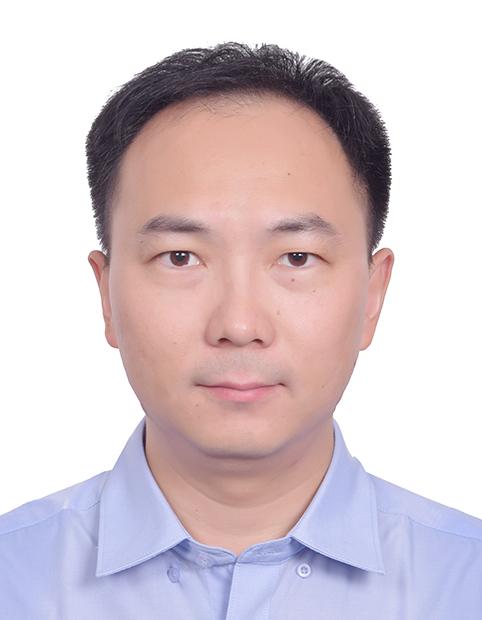}}]{Shimin Gong} (M'15) received the B.E. and M.E. degrees in Electronics and Information Engineering from Huazhong University of Science and Technology, Wuhan 430074, China, in 2008 and 2012, respectively, and the Ph.D. degree in Computer Engineering from Nanyang Technological University, Singapore, in 2014. He is currently a professor with the School of Intelligent Systems Engineering, Sun Yat-sen University, Shenzhen, China. His research interests include wireless powered communications, backscatter communications, and machine learning in wireless communications. He was the co-recipient of IEEE WCNC 2019 Best Paper Award on MAC and Cross-layer Design and 2023 IEEE Communications Society Best Survey Paper Award. He is an Associate Editor of IEEE Transactions on Vehicular Technology.
\vspace{-0.6cm}
\end{IEEEbiography}
\begin{IEEEbiography}[{\includegraphics[width=1in,height=1.25in,clip,keepaspectratio]{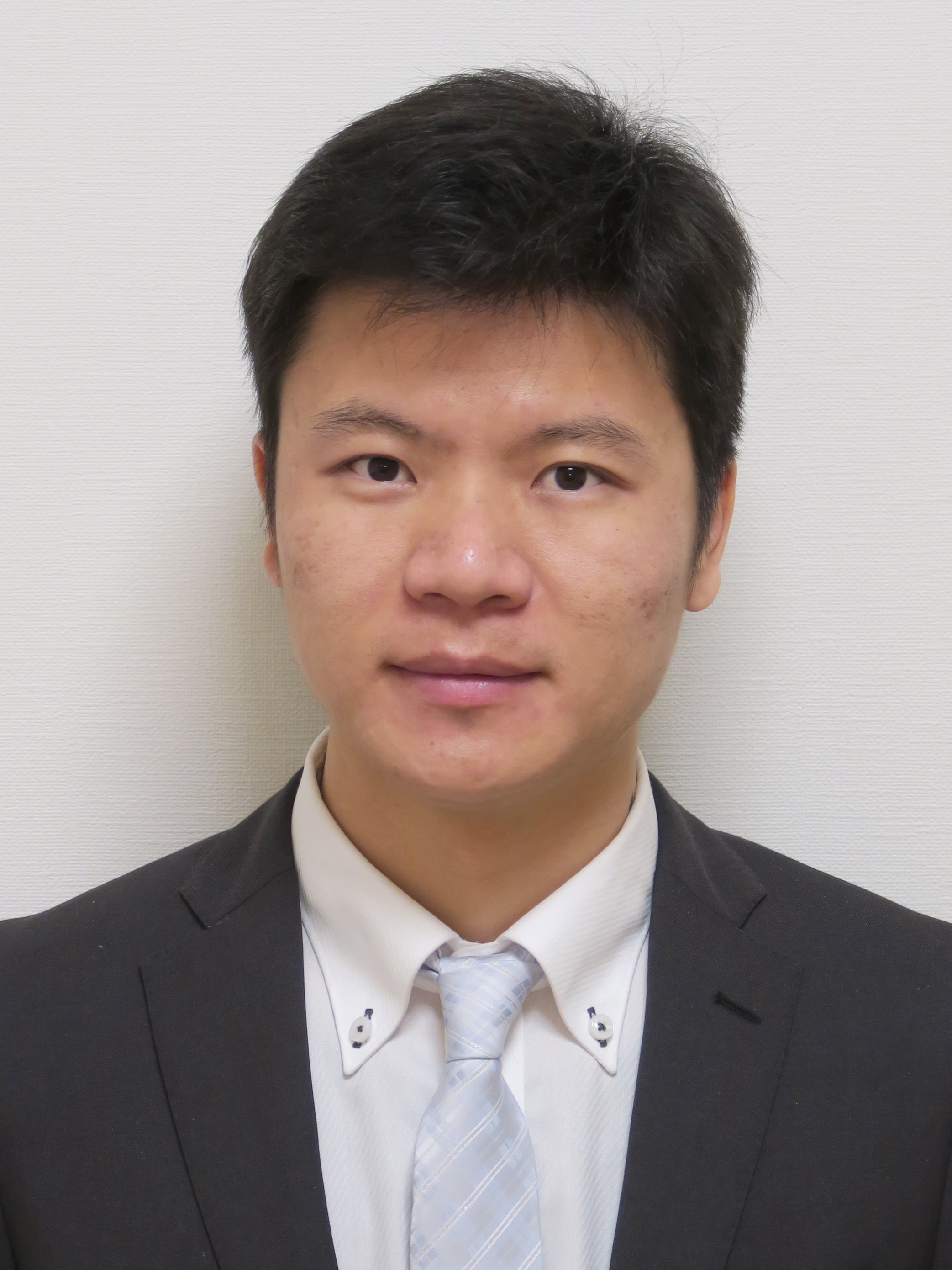}}]{Bo Gu (Member, IEEE)} received the Ph.D. degree from Waseda University, Tokyo, Japan, in 2013. He is currently a Professor with the School of Intelligent Systems Engineering, Sun Yat-sen University, Guangzhou, China. He was a Research Engineer with Sony Digital Network Applications, Yokohama shi, Japan, from 2007 to 2011, an Assistant Professor with Waseda University, from 2011 to 2016, and an Associate Professor with Kogakuin University, Tokyo, from 2016 to 2018. His research interests include Internet of Things, edge computing, network economics, and machine learning. He was the recipient of the IEEE ComSoc Communications Systems Integration and Modeling (CSIM) Technical Committee Best Journal Article Award in 2019, the Asia-Pacific Network Operations and Management Symposium (APNOMS) Best Paper Award in 2016, and the IEICE Young Researcher's Award in 2011. He is a member of IEICE.
\vspace{-0.6cm}
\end{IEEEbiography}
\begin{IEEEbiography}[{\includegraphics[width=1in,height=1.25in,clip,keepaspectratio]{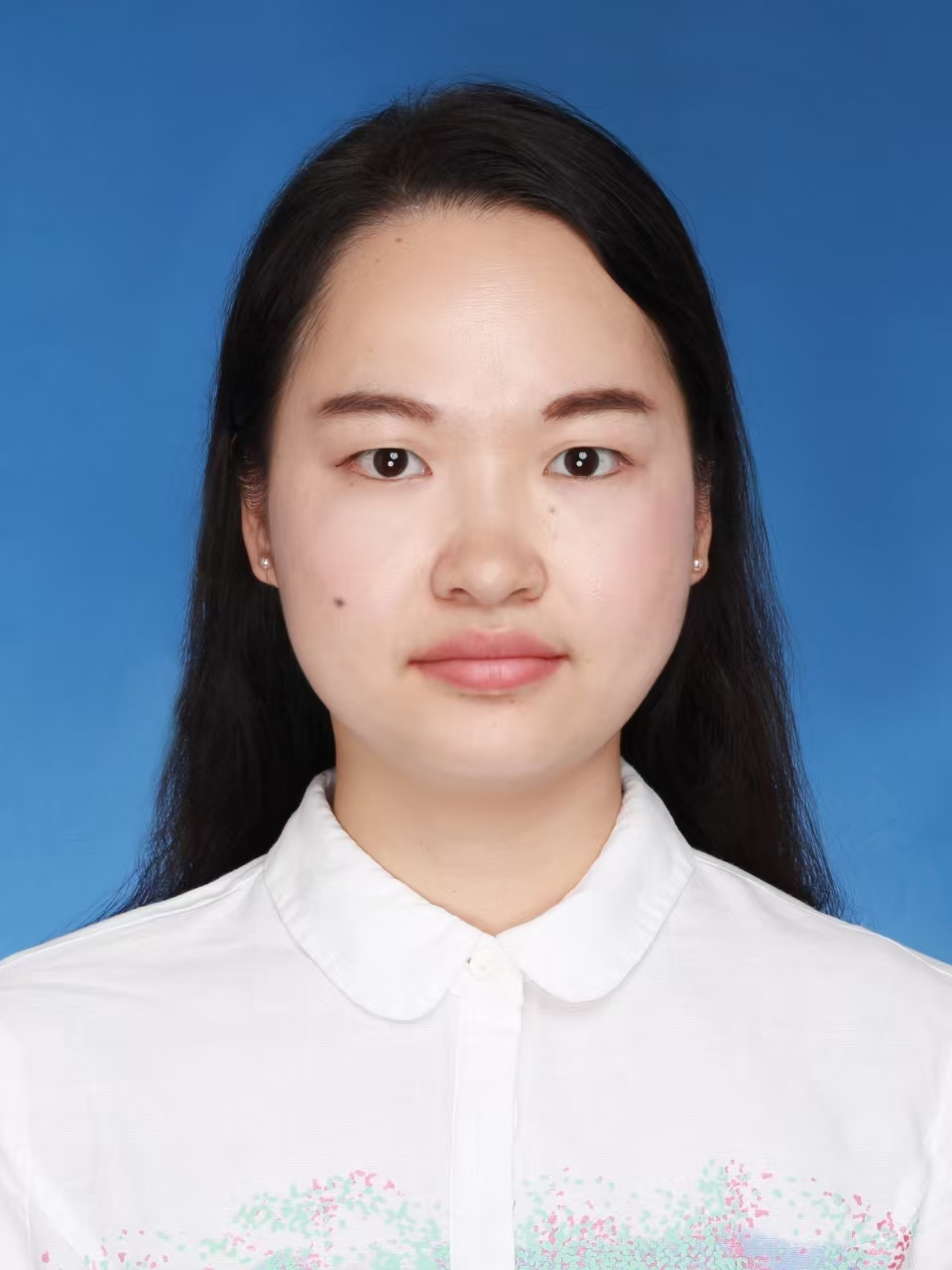}}]{Lanhua Li} is currently a Research Assistant Professor with the School of Intelligent Systems Engineering, Sun Yat-Sen University. She received the Ph.D. degree from the University of Chinese Academy of Sciences, Shenzhen, China, in 2021. Her research interests include wireless resource optimization, intelligent reflecting surface, backscatter communications, and UAV-enabled communication and sensing.
\vspace{-0.6cm}
\end{IEEEbiography}
\begin{IEEEbiography}[{\includegraphics[width=1in,height=1.25in,clip,keepaspectratio]{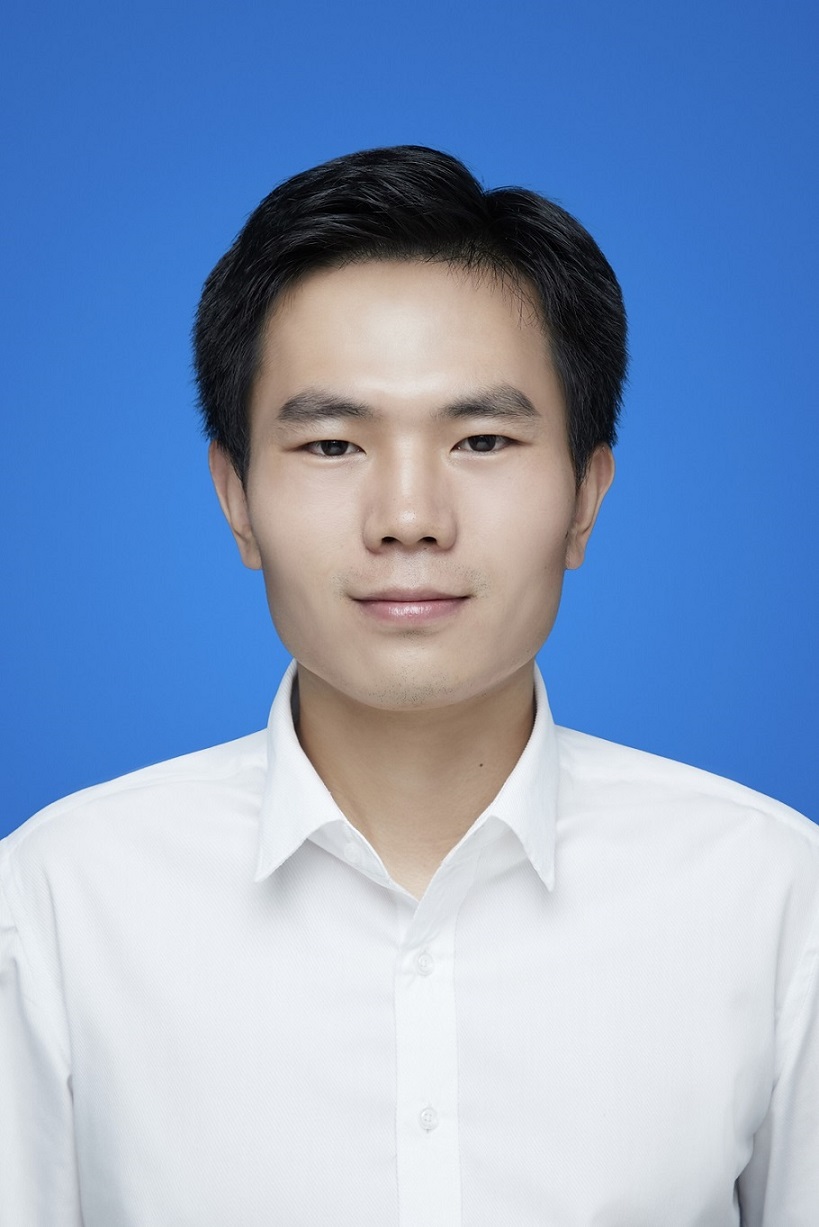}}]{Bin Lyu} (Senior Member, IEEE) received the B.E. and Ph.D. degrees from the Nanjing University of Posts and Telecommunications, Nanjing, China, in 2013 and 2018, respectively, where he is currently an Associate Professor. His research interests include wireless powered communications, backscatter communications, symbiotic communications, and mobile edge computing.
\vspace{-0.6cm}
\end{IEEEbiography}
\begin{IEEEbiography}[{\includegraphics[width=1in,height=1.25in,clip,keepaspectratio]{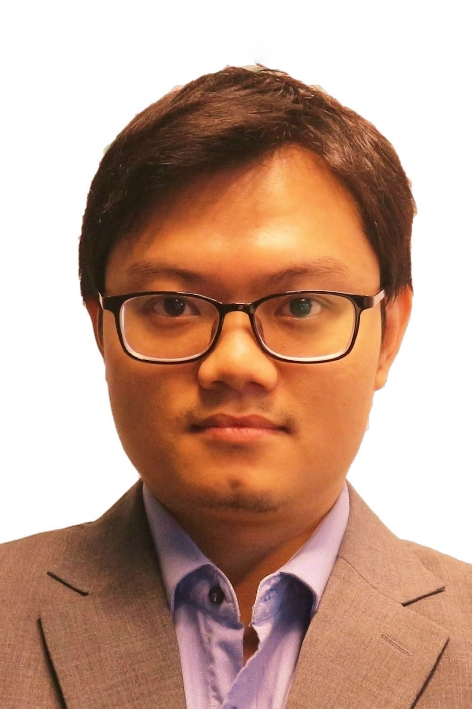}}]{Dinh Thai Hoang} (M'16, SM'22) is currently a faculty member at the School of Electrical and Data Engineering, University of Technology Sydney, Australia. He received his Ph.D. in Computer Science and Engineering from the Nanyang Technological University, Singapore 2016. His research interests include emerging wireless communications and networking topics, especially machine learning applications in networking, edge computing, and cybersecurity. He has received several precious awards, including the Australian Research Council Discovery Early Career Researcher Award, IEEE TCSC Award for Excellence in Scalable Computing for Contributions on ``Intelligent Mobile Edge Computing Systems" (Early Career Researcher), IEEE Asia-Pacific Board (APB) Outstanding Paper Award 2022, and IEEE Communications Society Best Survey Paper Award 2023. He is currently an Editor of IEEE TMC, IEEE TWC, IEEE TCOM and IEEE TNSE.
\vspace{-0.6cm}
\end{IEEEbiography}
\begin{IEEEbiography}[{\includegraphics[width=1in,height=1.25in,clip,keepaspectratio]{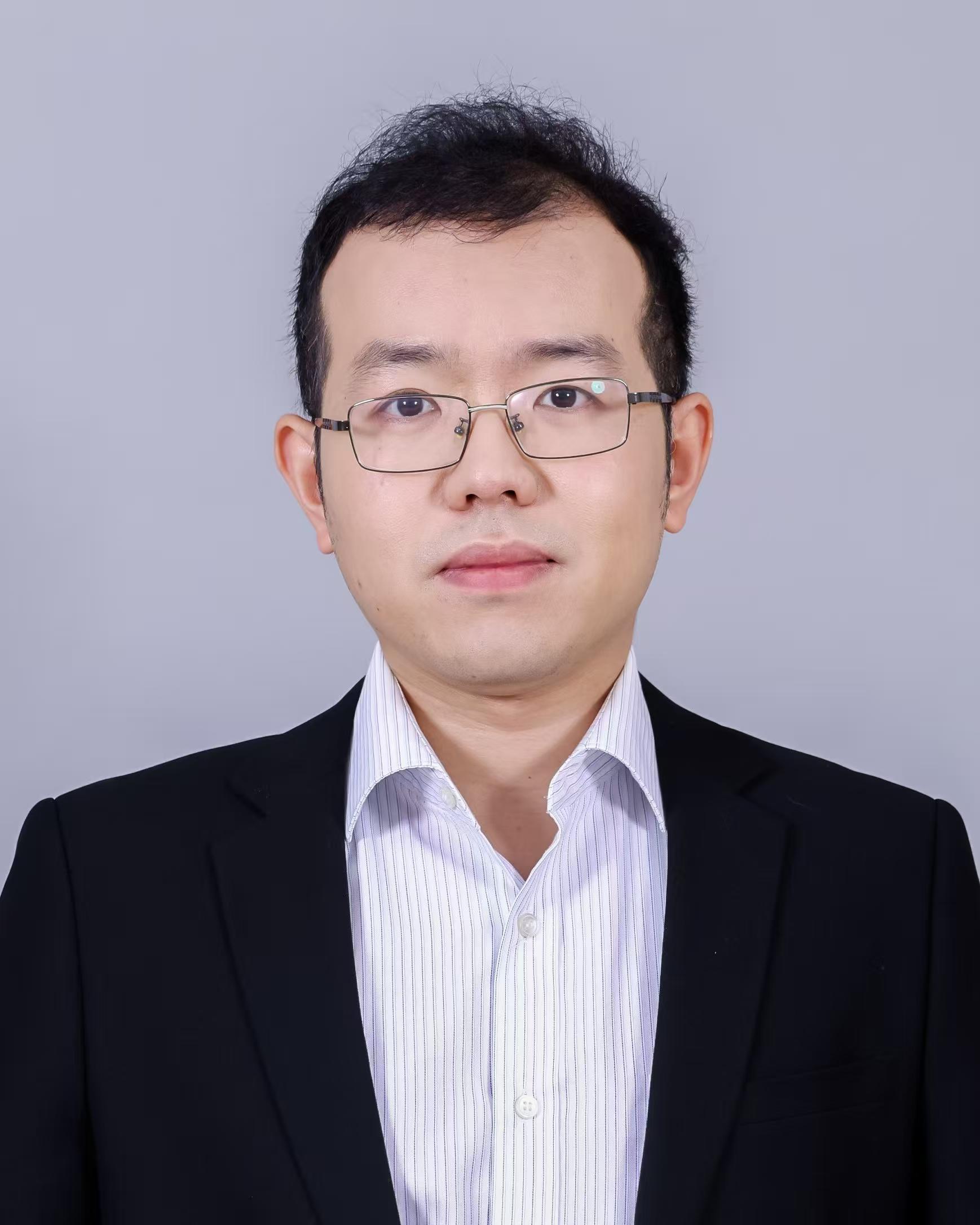}}]{Changyan Yi} (S'16-M'18-SM'24) received the Ph.D. degree in electrical and computer engineering from the University of Manitoba, MB, Canada, in 2018. He is currently a Professor with the College of Computer Science and Technology, Nanjing University of Aeronautics and Astronautics (NUAA), Nanjing, China. His research interests include stochastic optimization, game theory, incentive mechanism, queueing scheduling and machine learning with applications in resource management and decision making for various communication and networking systems.
\vspace{-0.6cm}
\end{IEEEbiography}
\end{document}